\documentclass[trackchanges, twocolumn]{aastex701}
\usepackage{graphicx}	% Including figure files
\usepackage{amsmath}	% Advanced maths commands
\usepackage{placeins}
\usepackage{hyperref}
\shorttitle{AGN-Driven Gas Motions in Hydra-A}
\shortauthors{A. Majumder et al.}
\accepted{\today}
\begin{document}

\title{Gas Motions in Hydra-A: XRISM Constraints on ICM Kinematics Across Jet-Inflated Cavities}

\author[orcid=0000-0002-3525-7186]{Anwesh Majumder}
\affiliation{Waterloo Centre for Astrophysics, Department of Physics and Astronomy, University of Waterloo, 200 University Avenue West, Waterloo,  Ontario N2L 3G1, Canada}
\affiliation{Space Research Organisation Netherlands, Niels Bohrweg 4, Leiden, South Holland 2333 CA, The Netherlands}
\email[show]{anwesh.majumder@uwaterloo.ca}  

\author[orcid=0000-0002-2622-2627]{B.R. McNamara}
\affiliation{Waterloo Centre for Astrophysics, Department of Physics and Astronomy, University of Waterloo, 200 University Avenue West, Waterloo,  Ontario N2L 3G1, Canada}
\email{mcnamra@uwaterloo.ca}

\author[orcid=0000-0003-0297-4493]{P.E.J. Nulsen}
\affiliation{Center for Astrophysics | Harvard \& Smithsonian, 60 Garden Street, Cambridge, MA 02138, USA}
\affiliation{ICRAR, University of Western Australia, 35 Stirling Hwy, Crawley, WA 6009, Australia}
\email{paulnulsen@gmail.com}

\author[orcid=0000-0001-5208-649X]{H. Russell}
\affiliation{School of Physics \& Astronomy, University of Nottingham, University Park, Nottingham NG7 2RD, UK}
\email{helen.russell@nottingham.ac.uk}

\author[orcid=0000-0002-8310-2218]{T. Rose}
\affiliation{Waterloo Centre for Astrophysics, Department of Physics and Astronomy, University of Waterloo, 200 University Avenue West, Waterloo,  Ontario N2L 3G1, Canada}
\email{thomas.rose@uwaterloo.ca}

\author[orcid=0000-0001-6670-6370]{T. Heckman}
\affiliation{Center for Astrophysical Sciences, William H. Miller III Department of Physics and Astronomy, Johns Hopkins University, Baltimore, MD 21218, USA}
\affiliation{School of Earth and Space Exploration, Arizona State University, Tempe, AZ 85287, USA}
\email{thechma1@jhu.edu}

\author[orcid=0000-0002-9714-3862]{A. Simionescu}
\affiliation{Space Research Organisation Netherlands, Niels Bohrweg 4, Leiden, South Holland 2333 CA, The Netherlands}
\affiliation{Leiden Observatory, Leiden University, Niels Bohrweg 2, Leiden, South Holland 2333 CA, The Netherlands}
\affiliation{Kavli Institute for the Physics and Mathematics of the Universe, The University of Tokyo, Kashiwa, Chiba 277-8583, Japan}
\email{a.simionescu@sron.nl}

\author[orcid=0000-0002-9378-4072]{A. Fabian}
\affiliation{Institute of Astronomy, University of Cambridge, Madingley Road, Cambridge CB3 0HA, UK}
\email{acf@ast.cam.ac.uk}

\author[orcid=0000-0002-6470-2285]{M.W. Wise}
\affiliation{Space Research Organisation Netherlands, Niels Bohrweg 4, Leiden, South Holland 2333 CA, The Netherlands}
\affiliation{Anton Pannekoek Institute, University of Amsterdam, Science Park 904, Amsterdam, North Holland 1098 XH, The Netherlands}
\email{m.w.wise@sron.nl}

\author[orcid=0000-0003-0392-0120]{N. Werner}
\affiliation{Department of Theoretical Physics and Astrophysics, Faculty of Science, Masaryk University, Kotlářská, Brno, 61137, Czech Republic}
\email{werner@physics.muni.cz}

\author[orcid=0000-0001-5226-8349]{M. McDonald}
\affiliation{MIT Kavli Institute for Astrophysics and Space Research, Massachusetts Institute of Technology, Cambridge, MA 02139, USA}
\email{mcdonald@space.mit.edu}

\begin{abstract}

We report on two deep XRISM observations of the central and northern regions of Hydra-A's X-ray atmosphere, covering the bubbles inflated by jets from the central galaxy's active galactic nucleus. We use spatial-spectral mixing, combining Chandra's high spatial resolution with XRISM's high spectral resolution, to investigate atmospheric kinematics. The velocity dispersion in the northern region, $\sigma_v = 140^{+30}_{-20}$ km s$^{-1}$, is comparable to that in the central region ($\sigma_v = 162 \pm 10$ km s$^{-1}$). We show that motion of the large-scale cocoon shock front could be responsible for the large dispersion toward the north. The dispersion in the northeast quarter of the central pointing, $\sigma_v = 260 \pm 50$ km s$^{-1}$, is among the highest measured. This region contains an X-ray-bright feature previously identified as metal-rich, possibly gas uplifted in the wake of previous-generation cavities. Dispersions in all other quarter regions are low ($\sigma_v \leq 180$ km s$^{-1}$) and consistent with previous XRISM results from other objects. The kinetic energy at the center is comparable to the enthalpies of the cavities, while in the north it is roughly an order of magnitude smaller. The jet thus drives gas motion efficiently at smaller scales ($r < 95$ kpc) and inefficiently at larger scales ($95-317$ kpc toward the north along the jet). A bulk flow toward our line of sight of $-100 \pm 30$ km s$^{-1}$ in the southwest quarter of the central pointing, measured with respect to the central galaxy, is also observed, possibly due to sloshing. No significant bulk motion was detected in any other region.

\end{abstract}


\section{Introduction} 

Jetted feedback from active galactic nuclei (AGN) is a critical mechanism that influences the evolution of the atmospheres around galaxies, groups, and clusters over cosmic time \citep{kondapally_cosmic_2023,heckman_global_2023,heckman_mergers_2024}. The Chandra X-ray Observatory has shown that jets provide enough energy to balance radiative cooling in X-ray atmospheres in the form of cavity enthalpy \citep{birzan_systematic_2004,rafferty_feedback-regulated_2006,panagoulia_volume-limited_2014}. However, the dominant mechanisms by which jet energy is thermalized are unknown. Shocks, sound wave heating (e.g., \citealt{fabian_deep_2003, nulsen_powerful_2005,randall_very_2015}, \citealt{Fabian2017}), mixing \citep{Hillel2016}, cosmic ray heating \citep{2008MNRAS.384..251G,2008A&A...484...51C,2017ApJ...844...13R,2019ApJ...871....6Y}, turbulence \citep{2005ApJ...622..205D,zhuravleva_turbulent_2014}, or a combination is plausible.

Over the past two years, the \textit{Resolve} microcalorimeter \citep{2022SPIE12181E..1SI} onboard the XRISM observatory \citep{2025PASJ...77S...1T} has been used to study gas motions in cluster atmospheres because of its superior spectral resolution ($\Delta E \sim 5$ eV at 6.4 keV). In most systems, gas motions in atmospheres lie below $200$ km s$^{-1}$ (e.g., \citealt{xrism_collaboration_disentangling_2025,xrism_collaboration_bulk_2025,xrism_collaboration_xrism_2025-1,fujita_xrism_2025,rose_xrism_2025}). M87 \citep{2026ApJ...998..210X} and Cygnus A \citep{2026ApJ...998..160M} are exceptions. Such low velocity dispersions, assumed to arise due to jet-induced turbulence, have raised questions about whether turbulence can balance cooling in clusters (\citealt{rose_xrism_2025,xrism_collaboration_bulk_2025,2026ApJ...998..210X,fujita_xrism_2025,li_simulating_2025,2026MNRAS.548ag775R,2026arXiv260419607M}). Moreover, even when turbulence can balance cooling, it does so only under optimistic assumptions where the velocity dispersion is ascribed purely to isotropic turbulence, as opposed to unresolved anisotropic bulk motion. Such complications arise because \textit{Resolve} has modest spatial resolution (half-power diameter of 1.3 arcmin at 6.4 keV; \citealt{2025PASJ...77S...1T}), which makes it impossible to resolve various substructures in the ICM from the underlying smooth gas. Furthermore, it can be difficult to isolate jet-induced turbulence from larger-scale turbulence due to mergers or sloshing (e.g., \citealt{2012A&A...544A.103V,2013ApJ...762...78Z,bourne_agn_2017,2025arXiv251212754B}). Therefore, further investigations are required to understand how velocity dispersions in different objects vary radially and azimuthally, both toward and away from known structures such as jets and cavities. Results from such investigations would show what fraction of the velocity dispersion observed by XRISM is jet-induced isotropic turbulence.

In this work, we discuss the well-studied Fanaroff-Riley type I (FR I; \citealt{1974MNRAS.167P..31F}) radio source Hydra-A \citep{Lane2004}. The radio jet emanating from the central AGN is powerful ($L = 2 \times 10^{45}$ erg s$^{-1}$; \citealt{nulsen_clusterscale_2005}) and has created three pairs of cavities with powers of $(2-7) \times 10^{44}$ erg s$^{-1}$ \citep{wise_xray_2007}. The innermost cavities (A and B) are at a distance of $\sim 25$ kpc to the north and south of the central AGN. A second pair of cavities (named C and D) lies at a distance of 100 kpc and 60 kpc to the north and south, respectively. A third, and the oldest, pair of cavities can be seen at a distance of 220 kpc and 100 kpc to the north and south (E and F), respectively \citep{wise_xray_2007}. We present an analysis of two \textit{Resolve} pointings, one encompassing the inner and the middle pair of cavities \citep{rose_xrism_2025}, and another encompassing the northern old cavity. We analyze sub-array spectra as well as the off-center pointing toward the north to investigate the radial and azimuthal distributions of velocity dispersion and bulk velocity.

\begin{figure}
	\includegraphics[width=\columnwidth]{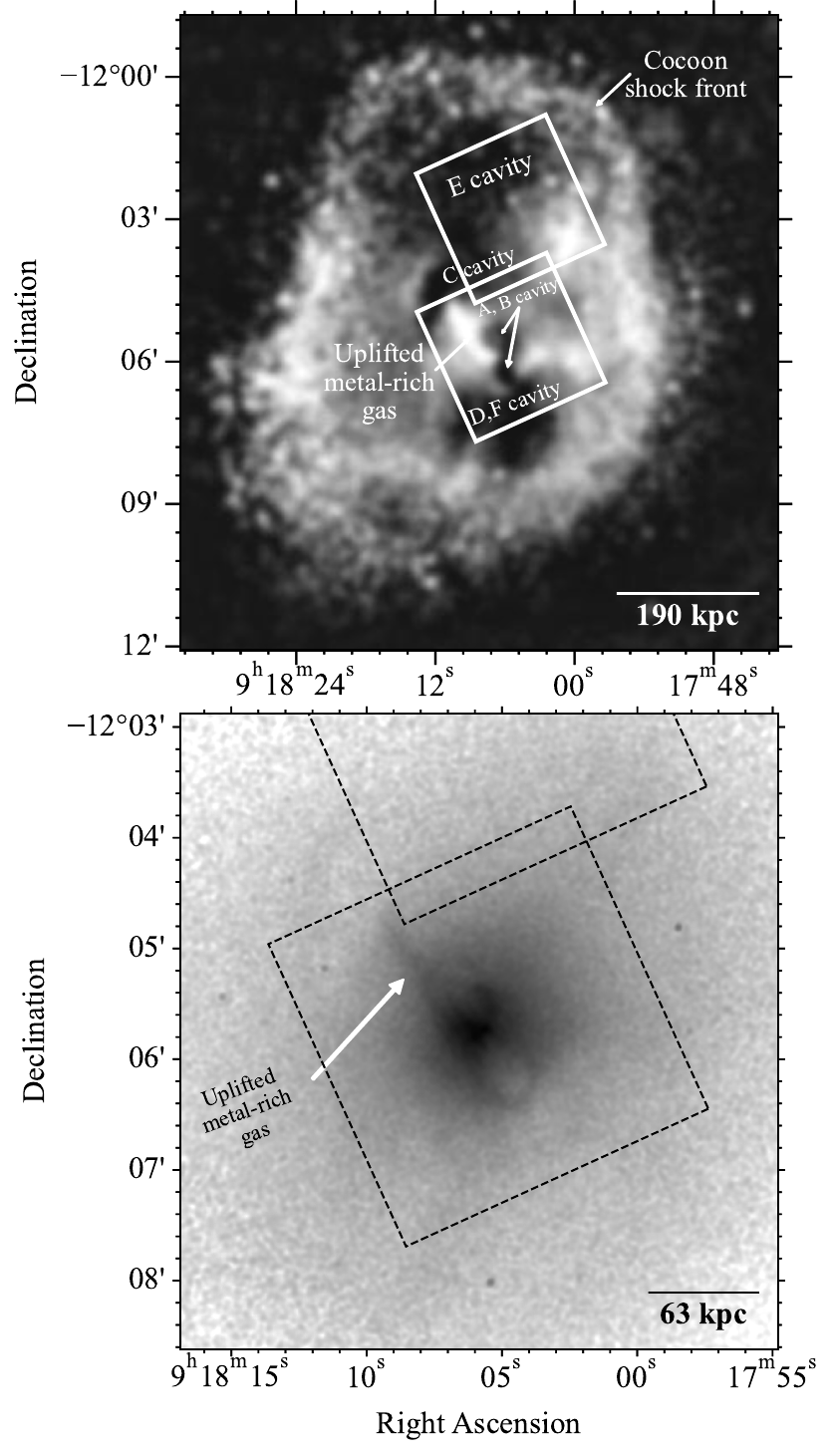}
    \caption{\emph{Top:} \textit{Chandra} X-ray residual image centered on the Hydra-A brightest cluster galaxy, from \citet{wise_xray_2007}. The authors subtracted a beta model fit to the cluster surface brightness profile, revealing multiple pairs of cavities visible out to 220 kpc. White rectangles show the fields of view of the two XRISM \textit{Resolve} observations analyzed in this paper. The labels mark various features in the image. \emph{Bottom:} A zoomed-in \textit{Chandra} image of the cluster containing Hydra-A, in the $0.5-2.0$ keV band. The image has been smoothed with a Gaussian kernel (radius of 5 pixels, $\sigma = 2.5$ pixels). The black dashed rectangles show the fields of view of the two XRISM \textit{Resolve} observations.}
    \label{fig:central_and_northern_pointings_outlines}
\end{figure}

In Section \ref{sec:data_reduction} we describe the data reduction procedure for XRISM, \textit{Chandra}, and MUSE. Sub-array analysis methods are described in Section \ref{sec:subarray_method}, and the spectral fitting method and modeling are described in Section \ref{sec:spec_fit_model}. Results are then presented in Section \ref{sec:ssm_results} followed by our interpretation in Section \ref{sec:discussion}.

We assume a $\Lambda$CDM cosmology with $H_0 = 70$ km s$^{-1}$ Mpc$^{-1}$, $\Omega_m = 0.3$, and $\Omega_{\Lambda} = 0.7$. The corresponding linear scale at the redshift of Hydra-A (MUSE stellar redshift, $z = 5.435 \times 10^{-2}$;  \citealt{Rose2019a}) is 63.4 kpc per arcmin. All errors reported in this work are of $1\sigma$ significance.

\begin{figure}
	\includegraphics[width=\columnwidth]{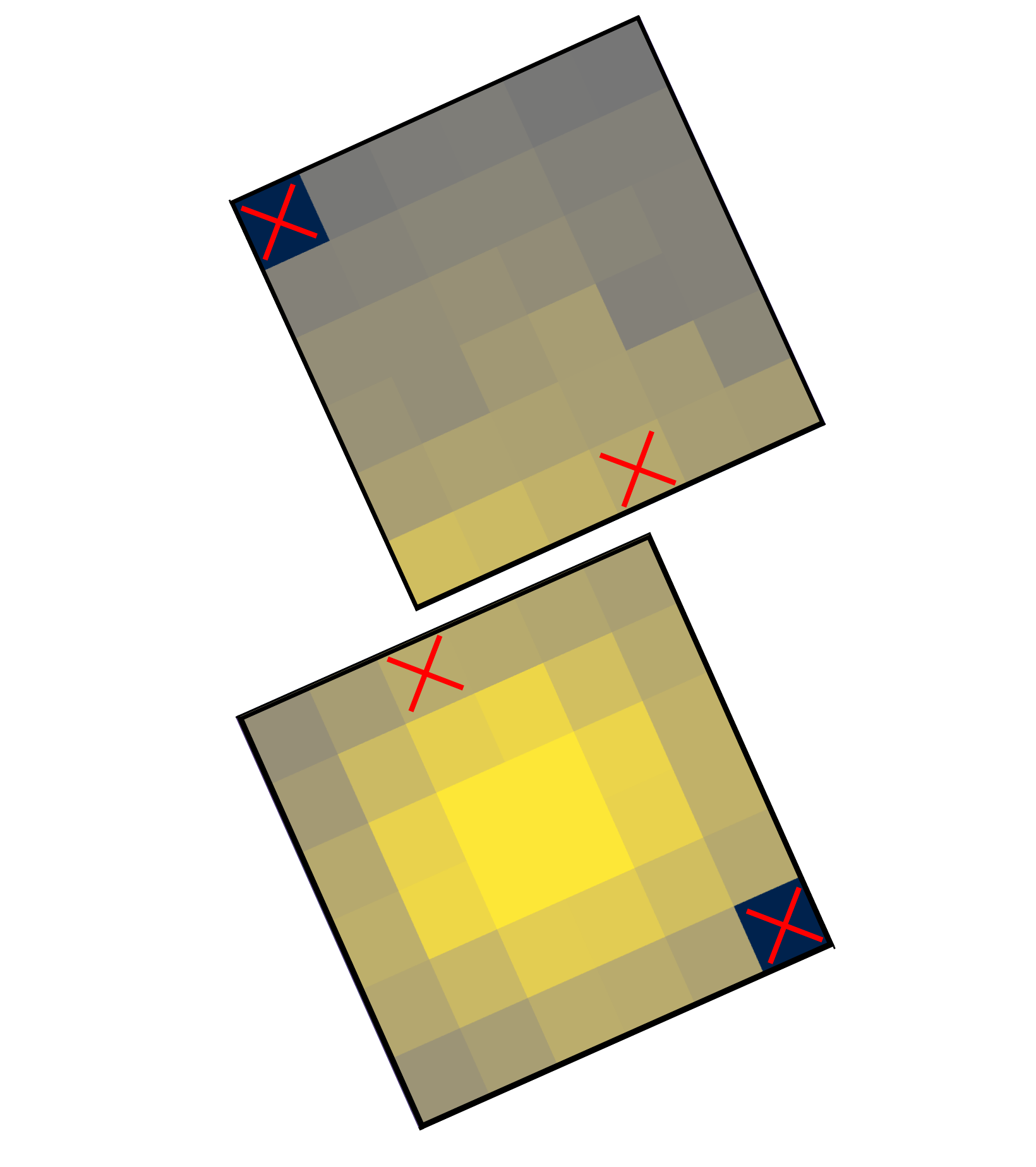}
    \caption{The two XRISM exposures presented in this paper. The two pointings are shown offset from their true sky positions for clarity. The lower of the two is centered on the brightest cluster galaxy, while the upper is aligned with the large, older northern cavity E (see Figure \ref{fig:central_and_northern_pointings_outlines}).}
    \label{fig:central_and_northern_pointings}
\end{figure}

\section{Observations and Data Reduction} \label{sec:data_reduction}

\subsection{XRISM}

In this work, we analyze two XRISM \textit{Resolve} observations of Hydra-A. The first observation (observation ID: 201070010) began on November 27, 2024, and was pointed at RA = $9^{\textrm{\scriptsize{h}}} 18^{\textrm{\scriptsize{m}}} 05^{\textrm{\scriptsize{s}}}.42$, Dec = $-12^{\circ} 05' 42''.34$, with a nominal aspect point roll of $114.49^{\circ}$ and centered on the brightest cluster galaxy (MCG-02-24-007, $z = 5.435 \times 10^{-2}$; \citealt{Rose2019a}). We will refer hereafter to this first pointing as the ``central pointing". A second observation (observation ID: 201071010) began on May 4, 2025, and was pointed at RA = $9^{\textrm{\scriptsize{h}}} 18^{\textrm{\scriptsize{m}}} 06^{\textrm{\scriptsize{s}}}.80$, Dec = $-12^{\circ} 02' 51''.18$ and had a nominal aspect point roll of $294.49^{\circ}$. This observation is thus centered further north with respect to the central pointing and covers the giant X-ray cavity previously seen with \textit{Chandra} data (cavity E in \citealt{wise_xray_2007}). We will refer to this second pointing as the ``northern pointing". The footprints covered by both observations are shown in Figure \ref{fig:central_and_northern_pointings_outlines}. 

All data were processed using HEASoft version 6.35.1 and \textit{Resolve} CALDB version 20250315. We filtered event files by applying the criterion \texttt{(((((RISE\_TIME+0.00075*DERIV\_MAX)>46)\
\&\&((RISE\_TIME+0.00075*DERIV\_MAX)<58))\&\&ITYPE<4)} \texttt{||(ITYPE==4))\&\&STATUS[4]==b0}. The calibration pixel 12, as well as pixel 27, was excluded from the analysis. Pixel 27 shows unexpected gain jumps not captured by the standard calibration. We include only high-resolution primary (Hp) events (highest energy resolution), which account for $\approx 94\%$ of the 1.7–12 keV events in the central pointing and $\approx 77\%$ in the northern pointing. After following these procedures, we obtained a total net exposure of 116.3 ks for the central pointing and 195.4 ks for the northern pointing.

For the central and northern pointings, the energy scale errors were $\sim 0.17$ eV\footnote{\href{https://heasarc.gsfc.nasa.gov/FTP/xrism/postlaunch/gainreports/2/201070010_resolve_energy_scale_report.pdf}{201070010\_resolve\_energy\_scale\_report.pdf}} and $\sim 0.1$ eV, respectively. Adding the current energy scale accuracy of $< 0.3$ eV\footnote{\href{https://tinyurl.com/2uh955jb}{https://heasarc.gsfc.nasa.gov/docs/xrism/proposals/POG/\\Resolve.html}} in quadrature results in a total bulk velocity systematic uncertainty of $\sim 18$ km s$^{-1}$ and $15$ km s$^{-1}$ at 6 keV for the central and northern pointings, respectively. The current line-spread function accuracy of 0.17 eV in the $6-7$ keV band results in a systematic uncertainty of $< 2$ km s$^{-1}$ for a $\sim$160 km s$^{-1}$ velocity dispersion.

XRISM had barycentric velocities of 27 km s$^{-1}$ (JD 2460639) and $-26$ km s$^{-1}$ (JD 2460800) during the central and northern pointing observations, respectively. The velocities were calculated using the equations in \citet{Wright2014}. The barycentric corrections were taken into account during spectral analysis (see Section \ref{subsec:spec_modelling} for more details). The low-Earth-orbit broadening is approximately 8 km s$^{-1}$, which, when added in quadrature with the true velocity broadening, results in a 0.2 km s$^{-1}$ increase in the observed $\sigma_v$.

\subsection{\textit{Chandra}}

We reprocessed all ACIS observations of the Hydra-A cluster in the \textit{Chandra} data archive, i.e., observation IDs 575, 576, 4969, and 4970. We used \texttt{CIAO 4.17.0} \citep{2006SPIE.6270E..1VF} and \texttt{CALDB 4.12.0} for reprocessing. The \texttt{CIAO} tool \texttt{chandra\_repro} was used to create level 2 event files with the option \texttt{check\_vf\_pha = yes}, providing additional cleaning of the particle background for any VFAINT mode observations. The event files were then cleaned by binning events in the $9-12$ keV energy range in 100 s intervals. Each count rate histogram was then fitted with a Gaussian function to determine the mean ($\mu$) and the standard deviation ($\sigma$). All events with a count rate greater than $1\sigma$ from the mean were rejected. We then focused on removing fast flares by binning the previously filtered events in the same energy range in 20 s intervals and repeating the Gaussian fits. All events with a count rate greater than $2\sigma$ from the mean were rejected, thus giving us a cleaned event file. 

All cleaned level 2 event files were then reprojected to a common tangent point with the tool \texttt{reproject\_obs}. Particle background event files for each reprojected event file were created by using the ``stowed" ACIS event files from \texttt{CALDB} and applying the appropriate gain correction. This ``stowed" particle background was measured with the instrument not exposed to the sky. We filtered these particle background events further by selecting \texttt{status=0} events for VFAINT mode observations. Background images from these events were then created in the $0.7-7.0$ keV band and combined to create a total background image for the field of view of all observation IDs.

Exposure-corrected images in the $0.7-7.0$ keV band for each event list were then created using the tool \texttt{flux\_obs} and combined. We then created a background-subtracted, exposure-corrected image for the entire field of view. This image was then used for selecting appropriate regions for raytracing and calculating suitable Ancillary Response Files (ARFs) for \textit{Resolve} spectra. 

\begin{figure*}
	\includegraphics[width=\textwidth, keepaspectratio]{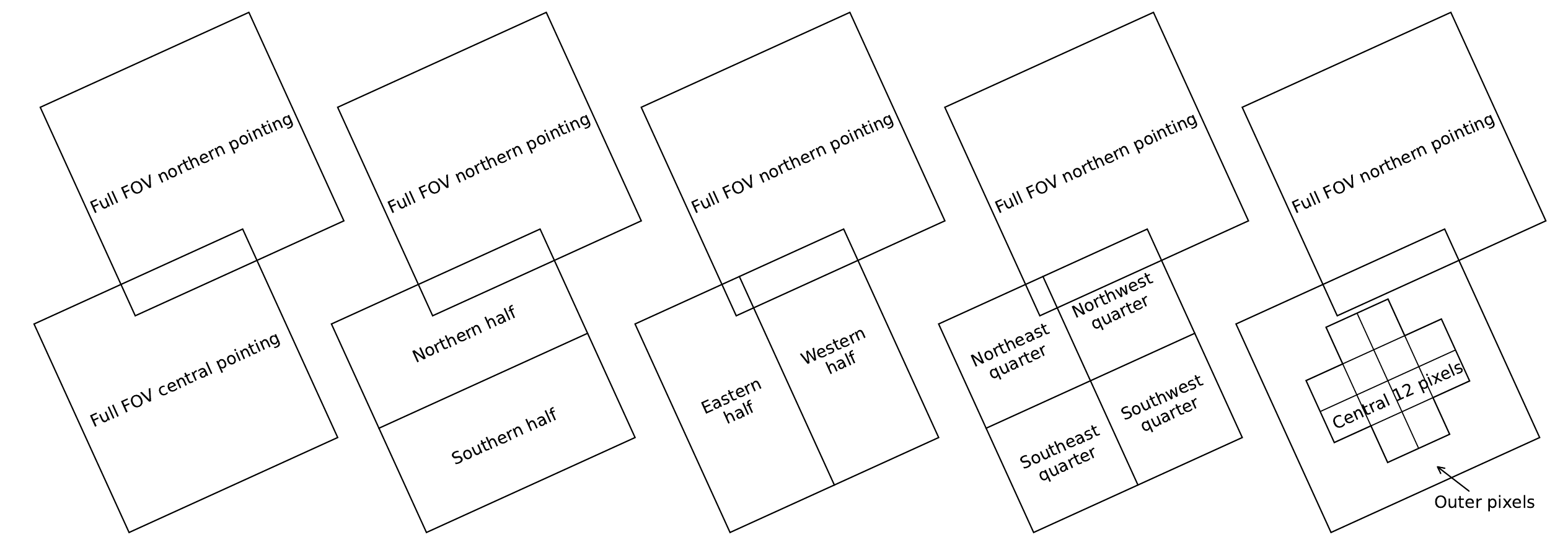}
    \caption{Schematic diagram of the two Resolve pointings and all SSM analysis regions. The central pointing is divided into halves, quarters, and central versus outer pixels, as described in Section \ref{sec:spec_fit_model}.}
    \label{fig:ssm_regions}
\end{figure*}

\subsection{MUSE} \label{subsec:MUSE}

A MUSE observation (PI: Stephen Hamer) centered on Hydra-A was performed on 2014 Dec 06, with an exposure time of 2700~s. The wavelength range covered by the observation is 475.0 to 935.2~nm, which includes the redshifted H$\alpha$ line. In Figure \ref{fig:MUSE_maps}, we show a map of the H$\alpha$-emitting gas, made by integrating the flux between 688.1 and 696.5~nm.

We created velocity and velocity dispersion maps by fitting multiple emission lines to the spectra from the MUSE data cube, including [S\,\textsc{ii}] ($\lambda = 671.832,\,673.271\,\mathrm{nm}$), H$\alpha$ ($656.461\,\mathrm{nm}$), [N\,\textsc{ii}] ($654.984,\,658.523\,\mathrm{nm}$), H$\beta$ ($486.269\,\mathrm{nm}$), [O\,\textsc{iii}] ($500.824\,\mathrm{nm}$), [S\,\textsc{iii}] ($907.110\,\mathrm{nm}$), and [O\,\textsc{i}] ($630.204\,\mathrm{nm}$).

For each spaxel that meets an integrated flux threshold of $3\sigma$, the spectrum is modeled as a sum of Gaussian components with a shared redshift $z$ and velocity dispersion $\sigma$, assuming a single kinematic component. To stabilize the fit, line amplitudes are fixed to physically motivated relative ratios, leaving four free parameters: overall normalization, redshift, dispersion, and a constant continuum offset. 

This stabilization may introduce small averaging errors but allows significantly fainter emission to be detected. We also perform the fits in order of decreasing signal strength, allowing previously fitted neighboring spaxels to provide initial guesses for $z$, which improves convergence and spatial coherence.

The fitted redshift is converted to a line-of-sight velocity via $v = cz$, with a constant offset applied for calibration, while the velocity dispersion is obtained from the Gaussian width as $\sigma_v = (\sigma / \lambda_{\mathrm{H}\alpha})\,c$.

Fits with unphysical dispersions (less than $4$ km s$^{-1}$ or greater than $400$ km\,s$^{-1}$, as judged by the appearance of the MUSE image) or that failed to converge are rejected. More extreme dispersion cutoffs introduced more noise in the MUSE images. For the chosen cutoffs, the resulting velocity and dispersion maps are shown in Figure \ref{fig:MUSE_maps}.

\section{Sub-array Analysis} \label{sec:subarray_method}

\subsection{Regions}

The central footprint was divided into halves, quarters, and the central 
twelve pixels versus the outer pixels. These regions are shown in Figure 
\ref{fig:ssm_regions}. Such choices were made to investigate gas motions in 
various regions of the core of the Hydra-A cluster. The count rate of 
$0.2299 \pm 0.0014$ cts s$^{-1}$ across the field of view is high enough to 
divide the field of view and still constrain the velocity dispersion. The whole field of view of the central pointing has previously been presented in \citet{rose_xrism_2025}. However, here we perform sub-array analysis in an 
attempt to constrain ICM kinematics in the radial and azimuthal directions.

The count rate for the full FOV northern pointing is only $0.0566\pm0.0005$ cts s$^{-1}$. This count rate is already similar to the quarter FOV central pointing count rates (see Table \ref{tab:count_rates}). We could not divide the FOV and perform SSM with these subdivided regions while still constraining the dispersions since SSM substantially enlarges error bars on dispersions (see Tables \ref{tab:cno_ICM_fits}, \ref{tab:nsno_ICM_fits}, \ref{tab:ewno_ICM_fits}, \ref{tab:nenwseswno_ICM_fits}, \ref{tab:12wo12no_ICM_fits}). Hence, we chose to use the whole field of view of the northern pointing in our analysis.

\begin{table}
    \centering
    \caption{Count rates for all our chosen regions in the 1.7-12.0 keV band}. 
    \begin{tabular}{cc}
    \hline
        Region & Counts s$^{-1}$\\
        \hline
        Full central pointing& $0.2299\pm0.0014$\\
        \hline
        Western halve & $0.1198\pm0.0010$\\
        Eastern halve & $0.1113\pm0.0010$\\
        Northern halve & $0.1168\pm0.0010$\\
        Southern halve & $0.1113\pm0.0010$\\
        \hline
        northwest quarter & $0.0636\pm0.0007$\\
        northeast quarter & $0.0532\pm0.0007$\\
        southwest quarter & $0.0560\pm0.0007$\\
        southeast quarter & $0.0581\pm0.0007$\\
        \hline
        Central 12 pixels& $0.1360\pm0.0011$\\
        Outer pixels& $0.0951\pm0.0009$\\
        \hline
        Full northern pointing & $0.0566\pm0.0005$ \\
        \hline
    \end{tabular}
    \tablecomments{The count rates exclude pixel 27, which has shown gain jumps and the calibration pixel 12.}
    \label{tab:count_rates}
\end{table}

\subsection{Spectral extraction}

Spectra were extracted from each region shown in Figure \ref{fig:ssm_regions}. The latest calibration file \texttt{xa\_rsl\_rmfparam\_20190101v006.fits} was then used to create an X-sized redistribution matrix file (RMF). The X-sized RMF models secondary response components best, at the expense of convolution speed. 

We also created a non-X-ray background (NXB) spectral file suitable for our purpose from a database of \textit{Resolve} night-Earth data using the task \texttt{rslnxbgen}. The standard recommended procedure to extract the NXB was followed.\footnote{\href{https://heasarc.gsfc.nasa.gov/docs/xrism/analysis/nxb/resolve_nxb_db.html}{https://heasarc.gsfc.nasa.gov/docs/xrism/analysis/nxb/\\resolve\_nxb\_db.html}} The NXB was modeled independently within the 1.7–17 keV energy band. This model included a power-law component along with Gaussian profiles for detector emission lines. The overall NXB normalization was first fitted for each spectrum, after which the individual emission line normalizations were adjusted. We then fixed this NXB model and used it in all subsequent spectral fits.

\section{Spectral fitting and modeling} \label{sec:spec_fit_model}

\subsection{Preparing Spectra for Fitting}

Spectral fits were computed using \texttt{SPEX v3.08.01} 
\citep{1996uxsa.conf..411K,2018zndo...2419563K,2020zndo...4384188K}.\footnote{\href{https://spex-xray.github.io/spex-help/index.html}{https://spex-xray.github.io/spex-help/index.html}} 
The extracted spectra, the RMF, ARFs, and the NXB suitable for each region 
were first converted to the SPEX format \citep{Kaastra2016}. All extracted 
response files from each region were optimally binned according to the 
instrumental energy resolution \citep{Kaastra2016} with the \texttt{SPEX} 
command \texttt{rbin} in the $1.7-12.0$ keV band. After optimal binning, we 
used the command \texttt{trafo} to arrange the responses that take into 
account contributions from different regions in multiple 
sectors.\footnote{\href{https://spex-xray.github.io/spex-help/theory/fitting/sectors.html}{https://spex-xray.github.io/spex-help/theory/fitting/sectors.html}} Photons in any given region could have (a) originated from that region, or (b) originated from other regions and scattered to the region of interest due to the broad point spread function of the mirror. Therefore, to fit spectra in any given region, we created multiple ARFs that account for the individual contributions of all regions where the photons could have originated. For example, to fit the spectrum from the northern half in the central pointing, we created an ARF that calculates the contribution from the northern half, a second ARF that calculates the contribution from the southern half of the central pointing, and a third ARF that calculates the contribution of the full FOV region of the northern pointing. The \texttt{HEASoft} task \texttt{xaarfgen} was used to calculate tailored ARFs for each case by raytracing from the combined \textit{Chandra} image of Hydra-A in the $0.7-7.0$ keV band. A million photons were randomly drawn for raytracing in each case, and a minimum limit of 100 photons was chosen for successful ARF creation. This analysis technique is commonly known as spatial-spectral mixing (SSM), which uses the high spatial resolution of the \textit{Chandra} data along with the high spectral resolution of the \textit{Resolve} data to maximize our understanding of gas properties. Below, we provide the details on how we loaded each of the created ARFs for each SSM analysis:

\begin{itemize}
    \item For the full FOV SSM analysis of the central and northern pointings, the ARF file accounting for photons from the full FOV of the central pointing was loaded in sector one, while that of the northern pointing was loaded in sector two. This process was repeated for both available spectra, i.e., the spectrum from the full FOV of the central pointing and that from the full FOV of the northern pointing.
    
    \item For the north–south sub-array SSM analysis of the central pointing, the ARF for the northern half was loaded in sector one, that for the southern half in sector two, and that for the full FOV of the northern pointing in sector three. This process was repeated for all three available spectra, i.e., the spectra from the northern and southern halves of the central pointing, and that from the full FOV of the northern pointing.

    \item For the east–west sub-array SSM analysis, we loaded the files from the eastern half in sector one, from the western half in sector two, and the full FOV files from the northern pointing in sector three. This process was repeated for all three available spectra, i.e., the spectra from the eastern and western halves of the central pointing, and that from the full FOV of the northern pointing.

    \item For the northeast, northwest, southeast, and southwest SSM analysis, we loaded files from the northeast quarter in sector one, the northwest quarter in sector two, the southeast quarter in sector three, the southwest quarter in sector four, and the full FOV files from the northern pointing in sector five. This process was repeated for all five available spectra, i.e., the spectra from the northeast, northwest, southeast, and southwest quarters of the central pointing, and that from the full FOV of the northern pointing.

    \item Finally, for the central-twelve-pixels versus outer-pixels SSM analysis, we loaded the files from the central twelve pixels in sector one, from the outer pixels in sector two, and the full FOV files from the northern pointing in sector three. This process was repeated for all three available spectra, i.e., the spectra from the central twelve pixels and the outer pixels of the central pointing, and that from the full FOV of the northern pointing.
    
\end{itemize}

\noindent After arranging the responses in the above manner, the spectra from each case were fitted simultaneously. This setup ensures that the spectral models of individual regions are folded through their appropriate ARFs and convolved to produce the total spectrum. Finally, the C-statistic was used in all our spectral fits for minimization \citep{1979ApJ...228..939C,2017A&A...605A..51K}. All abundances were measured with respect to the proto-solar abundance table of \citet{2009LanB...4B..712L}.

Although we report all our results by fitting in the $1.7-12.0$ keV band to maximize the spectral coverage for the central pointing, there are almost no source counts for the northern pointing at energies $<2$ keV and $>9$ keV. Therefore, we also rechecked all our analysis by fitting in the $2.0-8.0$ keV band. We found that the fitted parameter values changed under $1\sigma$ error bounds. Our results are thus insensitive to the exact energy band chosen for fitting purposes.

\subsection{Spectral Modelling} \label{subsec:spec_modelling}

The hot ICM was modeled with a single collisional ionization equilibrium (\texttt{cie}) component for all sub-array analyses. The normalization, temperature, and velocity broadening were varied independently. The $\alpha$-elements Si, S, Ar, and Ca were allowed to vary for fitting the spectrum from the full FOV of the central pointing but otherwise were coupled to Fe. Other elements from C to Ni were always coupled to the Fe abundance. The ion temperature was set equal to the electron temperature in all fits. The \texttt{cie} component was then redshifted using the \texttt{reds} model, whose redshift was left free in all spectral fits to measure any bulk velocity with respect to the central galaxy. Finally, the redshifted spectrum was absorbed through the Galactic hydrogen column of $n_H = 5.53 \times 10^{20}$ atoms cm$^{-2}$ using the \texttt{hot} model. The total weighted hydrogen absorption column was calculated using the method proposed by \citet{2013MNRAS.431..394W}.\footnote{\href{https://www.swift.ac.uk/analysis/nhtot/}{https://www.swift.ac.uk/analysis/nhtot/}} The temperature of the \texttt{hot} model was fixed to $10^{-6}$ keV for neutral absorption. Sky X-ray background was found to be negligible for our analysis and hence was not considered (See Section \ref{sec:skybkgsim}). The spectrum from the central pointing did not contain any high-energy power law tail, and hence no point source model was considered.

The barycentric velocities of the central and the northern pointings were unequal. To ensure that the models of the central pointing and the northern pointing were correct, we shifted any model that fits a spectrum from the central or northern pointing in energy space by the appropriate barycentric velocity. This shift was done using another \texttt{reds} model with the parameter \texttt{flag} set to 1 to ensure that SPEX calculates only a Doppler shift and excludes any time dilation due to a cosmological flow. The fitted redshift values from such a setup are then free from any effect due to the barycentric motion.

\begin{figure*}
	\includegraphics[width=\textwidth]{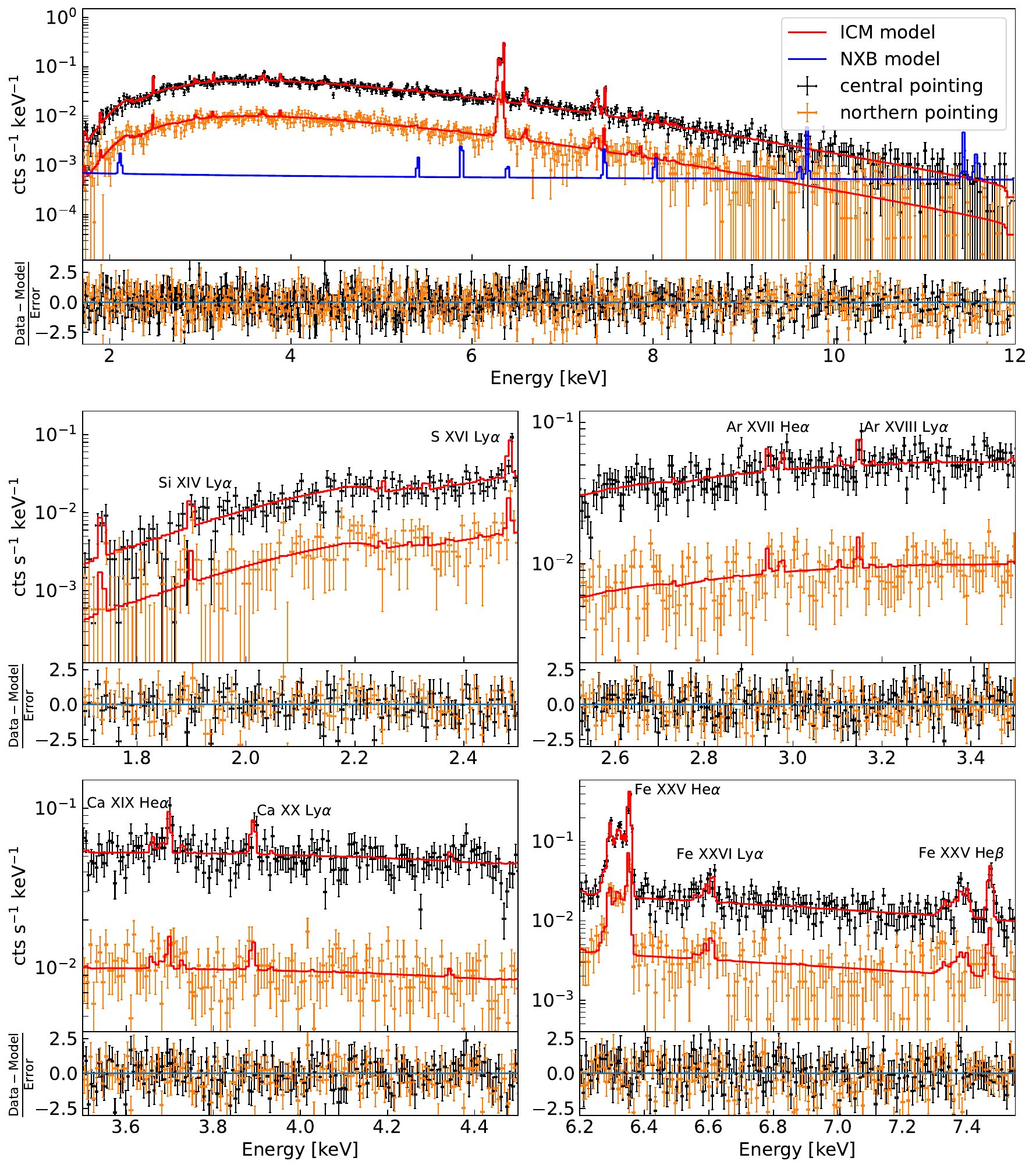}
    \caption{A single \texttt{cie} fit to the full FOV \textit{Resolve} spectrum 
for the central and northern pointings of Hydra-A. \emph{Top:} The ICM model 
fit to the $1.7-12.0$ keV band data along with the residuals. We also show 
the NXB model used for the fit. The spectrum has been binned by a factor of 5 
in this energy range for visual clarity. \emph{Middle left:} The same 
spectrum zoomed in on the $1.7-2.5$ keV band showing Si and S lines along 
with the residuals. \emph{Middle right:} The $2.5-3.5$ keV band showing Ar 
lines along with the residuals. \emph{Bottom left:} The $3.5-4.5$ keV band 
showing Ca lines along with the residuals. \emph{Bottom right:} The 
$6.2-7.6$ keV band showing Fe lines from the ICM. The middle and bottom 
panels have been binned by a factor of 2 for visual clarity.}
    \label{fig:cn_fit}
\end{figure*}

\section{SSM Results} \label{sec:ssm_results}

\subsection{Full FOV of the central pointing and full FOV of the northern pointing} \label{subsec:cenno_ssm}

One \texttt{cie} SSM fit to the full FOV of the central and the northern pointings is shown in Figure \ref{fig:cn_fit}. The fit parameters are reported in Table \ref{tab:cno_ICM_fits}. Using our SSM analysis, we recover the velocity dispersion of $\sim$160 km s$^{-1}$ previously measured for the central pointing by \citet{rose_xrism_2025}. The northern pointing shows a similar velocity dispersion of $140 \pm 30$ km s$^{-1}$. It contains the powerful E cavity ($P_{\textrm{cav}} = 6.8 \times 10^{44}$ erg s$^{-1}$; \citealt{wise_xray_2007}) and is also close to the cocoon shock front. These two observable features may together explain why the velocity dispersion does not decline with radius along the jet axis. We discuss the implications in 
Section \ref{sec:discussion}.

\begin{table}
    \centering
\caption{Best-fit parameters for the full FOV central-northern SSM in the $1.7-12.0$ keV band for Hydra-A.} 
\label{tab:cno_ICM_fits}
    \begin{tabular}{cc}
    \hline
    \hline
    Parameter&Value\\
    \hline
    \multicolumn{2}{c}{Central pointing} \\
         Redshift&  ($5.424 \pm 0.003) \times 10^{-2}$\\
         Norm ($n_e n_p V$)&$(17.3 \pm 0.3)\times 10^{66}$ cm$^{-3}$\\
        kT&$3.55 \pm 0.05$ keV\\
        $\sigma_{\textrm{1D}}$\tablenotemark{a} &$162 \pm 10$ km s$^{-1}$\\
        Si&$0.41^{+0.15}_{-0.13}$ $Z_{\odot}$\\
        S&$0.44 \pm 0.07$ $Z_{\odot}$\\
        Ar&$0.39 \pm 0.09$ $Z_{\odot}$\\
        Ca &$0.56 \pm 0.09$ $Z_{\odot}$\\
        Fe &$0.466 \pm 0.016$ $Z_{\odot}$\\
 Ni&$0.62 \pm 0.14$ $Z_{\odot}$\\
 \hline
 \multicolumn{2}{c}{Northern pointing}\\
 Redshift&($5.440 \pm 0.009) \times 10^{-2}$\\
 Norm ($n_e n_p V$)&$(2.92 \pm 0.12)\times 10^{66}$ cm$^{-3}$\\
 kT&$3.54 \pm 0.13$ keV\\
 $\sigma_{\textrm{1D}}$\tablenotemark{a} &$140 \pm 30$ km s$^{-1}$\\
 Fe &$0.37 \pm 0.03$ $Z_{\odot}$\\
 \hline
 C-stat (Expected C-stat)&$5277$ ($5336 \pm 99$)\\
 \hline
 \hline
    \end{tabular}

    \tablecomments{$n_e$ is the electron density, $n_p$ is the proton density and $V$ is the line-of-sight enclosed volume. All reported $\sigma_{1D}$ are velocity due to gas motions. The errors reported in this table are statistical.}

    \tablenotetext{a}{In SPEX \texttt{cie} models, $\sigma_{1D}$ is referred to as $v_{\textrm{\scriptsize{RMS}}}$.}
    
\end{table}

From the fitted redshifts, it is also possible to calculate the bulk motion of the ICM in the central and the northern pointings with respect to the central galaxy. The bulk velocity is

\begin{equation} \label{eq:bulk_velocity}
    v_{\textrm{\scriptsize{bulk, ICM-CG}}} = \frac{c (z_{\textrm{\scriptsize{ICM}}} - z_{\textrm{\scriptsize{CG}}})}{1 + z_{\textrm{\scriptsize{CG}}}},
\end{equation}

where $z_{\textrm{\scriptsize{CG}}} = (5.435 \pm 0.005) \times 10^{-2}$ is the MUSE optical redshift of the central galaxy \citep{Rose2019a} and $c$ is the speed of light. Applying this equation, we find an atmospheric bulk motion of $-32 \pm 17$ km s$^{-1}$ for the central pointing, with the negative sign indicating a blueshift with respect to the central galaxy. The northern pointing shows a bulk motion of $10 \pm 30$ km s$^{-1}$. The derived bulk motion values suggest that the atmosphere in both the central and northern pointings is nearly at rest with respect to the central galaxy. Our results are shown schematically in Figure \ref{fig:ssm_velocities}. We do not find evidence of resonance scattering with the current data (see Section \ref{sec:rsca}). Therefore, the Fe XXV-$w$ line was kept in all our subsequent analyses.

%\FloatBarrier

The Si/Fe, S/Fe, Ar/Fe, Ca/Fe, and Ni/Fe ratios for the ICM in the central 
pointing are $0.9 \pm 0.3$, $0.93 \pm 0.16$, $0.8 \pm 0.2$, $1.2 \pm 0.2$, and $1.3 \pm 0.3$, respectively. The results indicate solar ratios within errors. The Si/Fe, Ar/Fe, Ca/Fe, and Ni/Fe ratios are consistent (within $1\sigma$) with those of \citet{simionescu_chemical_2009}, derived from \textit{XMM-Newton} EPIC data. However, S/Fe is $1.9\sigma-2.7\sigma$ higher when compared with their $0'-0.5'$, $0.5'-1.0'$, and $1.0'-2.0'$ radius results. Our result thus clearly prefers proto-solar abundance ratios within \textit{Resolve}'s $3' \times 3'$ FOV for all elements, as opposed to the sub-solar S abundance ratio derived by \citet{simionescu_chemical_2009}.

We attempted a two-temperature \texttt{cie} model fit for the ICM in the central pointing. However, the reduction in \texttt{cstat} was insignificant (C-stat = 5274) for this more complex model. We also attempted to fit a Gaussian Differential Emission Measure (GDEM) model to our data. However, the fit improvement was again negligible (C-stat = 5278). \citet{simionescu_chemical_2009} found that the GDEM model fits X-ray gas properties better than a single-temperature model and is more physical compared to a two-temperature model, using the Fe-L complex from \textit{XMM-Newton} EPIC and RGS data. \textit{Resolve} is insensitive below 1.7 keV, and thus our fit is unable to find a multi-temperature distribution. Consequently, there may well be a multi-temperature distribution from unresolved cool gas blobs, as suggested by \citet{simionescu_chemical_2009}.

\subsection{Northern and southern halves of the central pointing and full FOV of the northern pointing}\label{subsec:nsno_ssm}

Next, we analyze the northern and southern halves of the central pointing along with the full northern pointing. The single-\texttt{cie} fit parameters are shown in Table \ref{tab:nsno_ICM_fits}. The fitted velocity dispersions indicate that the northern half of the central pointing has a significantly higher dispersion compared to the southern half. Without the northern half velocity dispersion of $240^{+40}_{-30}$ km s$^{-1}$, the full FOV velocity dispersion could be as low as $100^{+20}_{-30}$ km s$^{-1}$. In the subsequent sections, we pinpoint the location and the source driving this higher gas velocity. The velocity dispersion of the northern pointing ($\sigma_v = 130 \pm 30$ km s$^{-1}$) is similar to what was obtained in the full FOV SSM analysis.

\begin{table}
    \centering
\caption{Best-fit parameters for the northern, southern halves of central pointing and northern pointing full FOV SSM analysis in the $1.7-12.0$ keV band for Hydra-A.} 
\label{tab:nsno_ICM_fits}
    \begin{tabular}{cc}
    \hline
    \hline
    Parameter&Value\\
    \hline
    \multicolumn{2}{c}{Northern half of the central pointing} \\
         Redshift&  $5.447^{+0.008}_{-0.010} \times 10^{-2}$\\
         Norm ($n_e n_p V$)&$(8.7 \pm 0.3)\times 10^{66}$ cm$^{-3}$\\
        kT&$3.33^{+0.1}_{-0.09}$ keV\\
        $\sigma_{\textrm{1D}}$ &$240^{+40}_{-30}$ km s$^{-1}$\\
        Fe &$0.50 \pm 0.04$ $Z_{\odot}$\\
 \hline
 \multicolumn{2}{c}{Southern half of the central pointing} \\
         Redshift&  $5.412^{+0.006}_{-0.005} \times 10^{-2}$\\
         Norm ($n_e n_p V$)&$(8.5 \pm 0.3)\times 10^{66}$ cm$^{-3}$\\
        kT&$3.8 \pm 0.1$ keV\\
        $\sigma_{\textrm{1D}}$ &$100^{+20}_{-30}$ km s$^{-1}$\\
        Fe &$0.44 \pm 0.03$ $Z_{\odot}$\\
 \hline
 \multicolumn{2}{c}{Northern pointing}\\
 Redshift& $5.431^{+0.009}_{-0.008} \times 10^{-2}$\\
 Norm ($n_e n_p V$)&$ (2.95 \pm 0.13) \times 10^{66}$ cm$^{-3}$\\
 kT&$3.63^{+0.14}_{-0.13}$ keV\\
 $\sigma_{\textrm{1D}}$ &$130 \pm 30$ km s$^{-1}$\\
 Fe &$0.36 \pm 0.04$ $Z_{\odot}$\\
 \hline
 C-stat (Expected C-stat)&$7721$ ($7783 \pm 118$)\\
 \hline
 \hline
\end{tabular}
\end{table}

The bulk velocity of each region was calculated according to its redshift with respect to the central galaxy using Equation \ref{eq:bulk_velocity}. For the northern half, we obtain a bulk velocity of $30 \pm 30$ km s$^{-1}$, while the southern half has a bulk motion of $-70 \pm 20$ km s$^{-1}$.Thus, the northern half is essentially at rest, while the southern half has a bulk motion toward our line of sight ($3.5\sigma$ confidence). Finally, the redshift of the northern pointing indicates a bulk motion of $-10 \pm 30$ km s$^{-1}$.

\begin{table}
    \centering
\caption{Best-fit parameters for the eastern, western halves of central pointing and northern pointing full FOV SSM analysis in the $1.7-12.0$ keV band for Hydra-A.} 
\label{tab:ewno_ICM_fits}
    \begin{tabular}{cc}
    \hline
    \hline
    Parameter&Value\\
    \hline
    \multicolumn{2}{c}{Eastern half of the central pointing} \\
         Redshift&  $5.419^{+0.008}_{-0.007} \times 10^{-2}$\\
         Norm ($n_e n_p V$)&$(10.5 \pm 0.3)\times 10^{66}$ cm$^{-3}$\\
        kT&$3.39 \pm 0.08$ keV\\
        $\sigma_{\textrm{1D}}$ &$180^{+30}_{-40}$ km s$^{-1}$\\
        Fe &$0.43 \pm 0.03$ $Z_{\odot}$\\
 \hline
 \multicolumn{2}{c}{Western half of the central pointing} \\
         Redshift&  $(5.432 \pm 0.008) \times 10^{-2}$\\
         Norm ($n_e n_p V$)&$(6.7 \pm 0.3)\times 10^{66}$ cm$^{-3}$\\
        kT&$3.81^{+0.14}_{-0.12}$ keV\\
        $\sigma_{\textrm{1D}}$ &$150^{+50}_{-30}$ km s$^{-1}$\\
        Fe &$0.53^{+0.05}_{-0.04}$ $Z_{\odot}$\\
 \hline
 \multicolumn{2}{c}{Northern pointing}\\
 Redshift& $(5.44 \pm 0.01) \times 10^{-2}$\\
 Norm ($n_e n_p V$)&$ (3.08 \pm 0.13) \times 10^{66}$ cm$^{-3}$\\
 kT&$3.48^{+0.13}_{-0.12}$ keV\\
 $\sigma_{\textrm{1D}}$ &$140 \pm 30$ km s$^{-1}$\\
 Fe &$0.36 \pm 0.04$ $Z_{\odot}$\\
 \hline
 C-stat (Expected C-stat)&$7630$ ($7730 \pm 118$)\\
 \hline
 \hline
\end{tabular}
\end{table}

\begin{table}
    \centering
\caption{Best-fit parameters for the northeast, northwest, southeast, southwest quarters of central pointing and northern pointing full FOV SSM analysis in the $1.7-12.0$ keV band for Hydra-A.} 
\label{tab:nenwseswno_ICM_fits}
    \begin{tabular}{cc}
    \hline
    \hline
    Parameter&Value\\
    \hline
    \multicolumn{2}{c}{Northeast quarter of the central pointing} \\
         Redshift&  $(5.42 \pm 0.02) \times 10^{-2}$\\
         Norm ($n_e n_p V$)&$(5.0 \pm 0.3)\times 10^{66}$ cm$^{-3}$\\
        kT&$3.21^{+0.16}_{-0.15}$ keV\\
        $\sigma_{\textrm{1D}}$ &$260 \pm 50$ km s$^{-1}$\\
        Fe &$0.49 \pm 0.07$ $Z_{\odot}$\\
 \hline
 \multicolumn{2}{c}{Northwest quarter of the central pointing} \\
         Redshift&  $5.471^{+0.016}_{-0.013} \times 10^{-2}$\\
         Norm ($n_e n_p V$)&$(3.6 \pm 0.3)\times 10^{66}$ cm$^{-3}$\\
        kT&$3.5 \pm 0.2$ keV\\
        $\sigma_{\textrm{1D}}$ &$120^{+60}_{-50}$ km s$^{-1}$\\
        Fe &$0.54^{+0.09}_{-0.08}$ $Z_{\odot}$\\
 \hline
\multicolumn{2}{c}{Southeast quarter of the central pointing} \\
         Redshift&  $(5.419 \pm 0.009) \times 10^{-2}$\\
         Norm ($n_e n_p V$)&$(5.3 \pm 0.3)\times 10^{66}$ cm$^{-3}$\\
        kT&$3.57 \pm 0.16$ keV\\
        $\sigma_{\textrm{1D}}$ &$80 \pm 40$ km s$^{-1}$\\
        Fe &$0.39 \pm 0.05$ $Z_{\odot}$\\
 \hline
\multicolumn{2}{c}{Southwest quarter of the central pointing} \\
         Redshift&  $5.400^{+0.011}_{-0.009} \times 10^{-2}$\\
         Norm ($n_e n_p V$)&$(3.1 \pm 0.2)\times 10^{66}$ cm$^{-3}$\\
        kT&$4.2^{+0.3}_{-0.2}$ keV\\
        $\sigma_{\textrm{1D}}$ &$110^{+40}_{-60}$ km s$^{-1}$\\
        Fe &$0.53 \pm 0.08$ $Z_{\odot}$\\
 \hline
 \multicolumn{2}{c}{Northern pointing}\\
 Redshift& $(5.425 \pm 0.010) \times 10^{-2}$\\
 Norm ($n_e n_p V$)&$ (3.09 \pm 0.14) \times 10^{66}$ cm$^{-3}$\\
 kT&$3.58^{+0.15}_{-0.13}$ keV\\
 $\sigma_{\textrm{1D}}$ &$140 \pm 30$ km s$^{-1}$\\
 Fe &$0.36 \pm 0.04$ $Z_{\odot}$\\
 \hline
 C-stat (Expected C-stat)&$12014$ ($12165 \pm 145$)\\
 \hline
 \hline
\end{tabular}
\end{table}

%\FloatBarrier

\subsection{Eastern and western halves of the central pointing and full FOV of the northern pointing} \label{subsec:ewno_ssm}

We repeat our SSM analysis for the eastern and western halves of the central pointing along with the full northern pointing. The fit parameters are shown in Table \ref{tab:ewno_ICM_fits}. We observe similar velocity dispersions of $180^{+30}_{-40}$ km s$^{-1}$ and $150_{-30}^{+40}$ km s$^{-1}$ in the eastern and western halves, respectively . The mean velocity dispersion in the eastern half is higher, although the difference is not statistically significant. When combined with the results from Section \ref{subsec:nsno_ssm}, however, this result indicates that the velocity dispersion in the northeast of Hydra-A may be high. We show in Section \ref{subsec:nenwseswno_ssm} that this is indeed the case.

The bulk velocity with respect to the central galaxy can similarly be calculated as in previous subsections. For the eastern half, we obtain a bulk velocity of $-50 \pm 30$ km s$^{-1}$, while for the western half, we calculate a bulk velocity of $-10 \pm 30$ km s$^{-1}$. Thus, the ICM in both halves is at or close to rest with respect to the central galaxy. Furthermore, the redshift of the northern pointing suggests a bulk motion of $10 \pm 30$ km s$^{-1}$.

\begin{table*}
    \centering
\caption{Best-fit parameters for the central twelve pixels, outer pixels of the central pointing and northern pointing full FOV SSM analysis in the $1.7-12.0$ keV band for Hydra-A. The parameters without errors were kept fixed during the fitting.} 
\label{tab:12wo12no_ICM_fits}
    \begin{tabular}{ccc}
    \hline
    \hline
    Parameter   &   Value   &   Value\\
            & (all parameters free)     & (with frozen parameters) \\
    \hline
    \multicolumn{3}{c}{Central twelve pixels of the central pointing} \\
         Redshift&  $(5.432 \pm 0.005) \times 10^{-2}$&  $(5.432 \pm 0.005) \times 10^{-2}$\\
         Norm ($n_e n_p V$)&$(11.7 \pm 0.3)\times 10^{66}$ cm$^{-3}$&$(11.7 \pm 0.3)\times 10^{66}$ cm$^{-3}$\\
        kT&$3.52 \pm 0.08$ keV&$3.52^{+0.08}_{-0.07}$ keV\\
        $\sigma_{\textrm{1D}}$ &$160^{+40}_{-30}$ km s$^{-1}$&$160$ km s$^{-1}$\\
        Fe &$0.51 \pm 0.03$ $Z_{\odot}$&$0.52 \pm 0.03$ $Z_{\odot}$\\
 \hline
 \multicolumn{3}{c}{Outer pixels of the central pointing} \\
         Redshift&  $5.396^{+0.014}_{-0.023} \times 10^{-2}$&  $5.397^{+0.014}_{-0.012} \times 10^{-2}$\\
         Norm ($n_e n_p V$)&$(5.6 \pm 0.4)\times 10^{66}$ cm$^{-3}$&$(5.6 \pm 0.4)\times 10^{66}$ cm$^{-3}$\\
        kT&$3.7 \pm 0.2$ keV&$3.7 \pm 0.2$ keV\\
        $\sigma_{\textrm{1D}}$ &$130 \pm 130$ km s$^{-1}$&$130^{+50}_{-60}$ km s$^{-1}$\\
        Fe &$0.34 \pm 0.06$ $Z_{\odot}$&$0.33^{+0.06}_{-0.05}$ $Z_{\odot}$\\
 \hline
 \multicolumn{3}{c}{Northern pointing}\\
 Redshift& $(5.44 \pm 0.01) \times 10^{-2}$& $5.44 \times 10^{-2}$\\
 Norm ($n_e n_p V$)&$ (2.90 \pm 0.14) \times 10^{66}$ cm$^{-3}$&$2.9 \times 10^{66}$ cm$^{-3}$\\
 kT&$3.50^{+0.15}_{-0.14}$ keV&$3.5$ keV\\
 $\sigma_{\textrm{1D}}$ &$140^{+30}_{-20}$ km s$^{-1}$&$140$ km s$^{-1}$\\
 Fe &$0.40 \pm 0.04$ $Z_{\odot}$&$0.4$ $Z_{\odot}$\\
 \hline
 C-stat (Expected C-stat) & \multicolumn{2}{c}{$7636$ ($7738 \pm 118$)}\\
 \hline
 \hline
\end{tabular}
\end{table*}

%\FloatBarrier

\subsection{Northeast, northwest, southeast, southwest quarters of the central pointing and the full FOV of the northern pointing} \label{subsec:nenwseswno_ssm}

We next repeat our analysis by dividing the central pointing into quarters and analyzing them together with the northern pointing. The fit parameters are listed in Table \ref{tab:nenwseswno_ICM_fits}. A high velocity dispersion of $260 \pm 50$ km s$^{-1}$ is seen in the northeast quarter. Such a high dispersion has only been reported in the full FOV of Cygnus A \citep{2026ApJ...998..160M} and the Virgo cluster \citep{2026ApJ...998..210X}, although our uncertainty is much larger. The dispersions in the other quarters are low ($< 120$ km s$^{-1}$), with the southeast quarter reaching a dispersion as low as 80 km s$^{-1}$.

The northeast and southeast quarters have bulk velocities of $-40 \pm 60$ km s$^{-1}$ and $-50 \pm 30$ km s$^{-1}$, respectively. The ICM in these quarters is thus at or near rest with respect to the central galaxy. This is consistent with the near-zero ICM motion we calculated for the eastern half in Section \ref{subsec:ewno_ssm}. However, the northwest quarter has a bulk velocity of $100 \pm 40$ km s$^{-1}$, while the southwest quarter has a bulk velocity of $-100 \pm 30$ km s$^{-1}$. These two opposite bulk flows average out to the near-zero ICM motion we previously calculated for the western half in Section \ref{subsec:ewno_ssm}. We again discuss the implications of this result in Section \ref{sec:discussion}. The bulk motion of the northern pointing is constrained to $-30 \pm 30$ km s$^{-1}$.

\begin{figure*}
	\includegraphics[width=\textwidth, keepaspectratio]{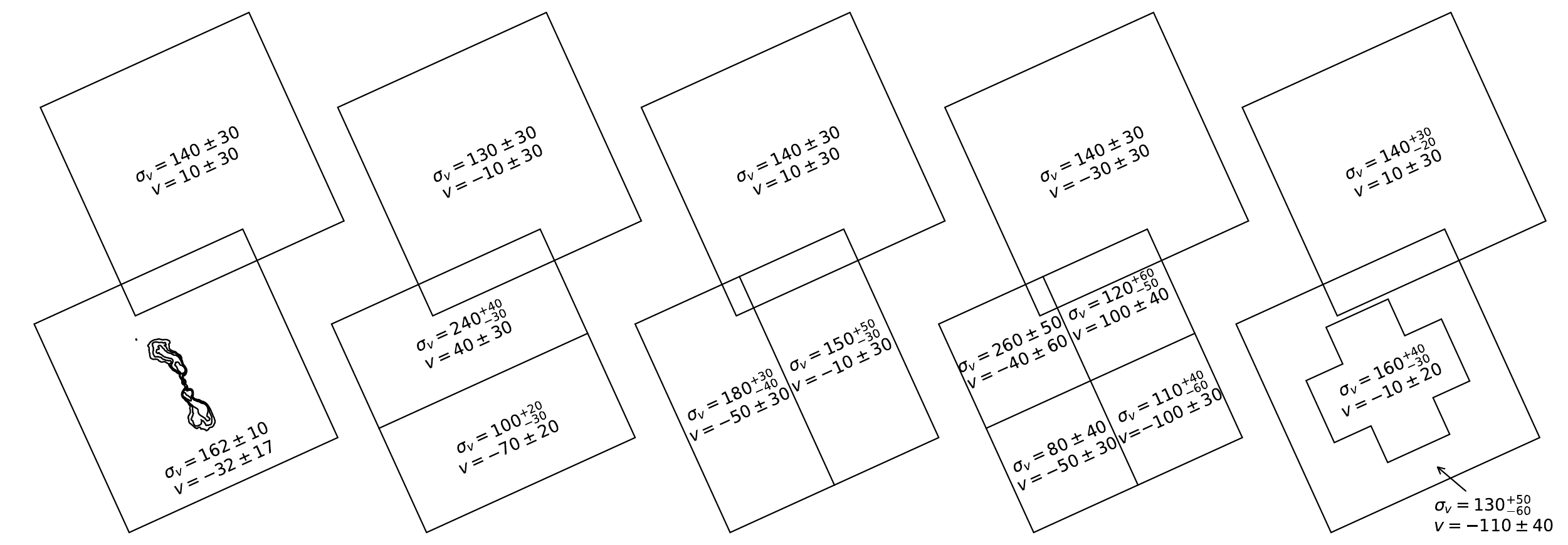}
    \caption{Same as Figure \ref{fig:ssm_regions} but with velocity dispersions, $\sigma_v$, and the bulk velocity with respect to the central galaxy, $v$, 
indicated for each region. All values are in units of km s$^{-1}$. The leftmost panel also shows the 5 GHz EVLA (Project 13B-088) radio lobes overlaid on the central pointing for reference.}
    \label{fig:ssm_velocities}
\end{figure*}

%\clearpage

\subsection{Central twelve pixels, outer pixels of the central pointing and full FOV of the northern pointing} \label{subsec:radial_profile}

Finally, we divide the central pointing into ``central twelve pixels'' and ``outer pixels'' and analyze them together with the northern pointing. The extraction regions for the central twelve pixels and outer pixels are shown in Figure \ref{fig:ssm_regions}. By dividing the central pointing this way, we can measure how the dispersion and bulk velocity vary radially in Hydra-A. The SSM fit parameters are noted in Table \ref{tab:12wo12no_ICM_fits}. We find that the central twelve pixels have a velocity dispersion of $160^{+40}_{-30}$ km s$^{-1}$ and the outer pixels have a dispersion of $130 \pm 130$ km s$^{-1}$. Thus, when all parameters are allowed to vary, we are unable to constrain the dispersion in the outer pixels of the central pointing due to the low count rate. We can, however, obtain a constraint by fixing all the ICM parameters in the northern pointing. The ICM parameters of the northern pointing are consistent across all SSM analyses, and therefore we can fix them without introducing additional assumptions. Furthermore, we also fix the dispersion of the central twelve pixels to the mean value (i.e., 160 km s$^{-1}$). This way, only the dispersion of the outer pixels from the central pointing is allowed to vary in our SSM analysis. When the analysis is performed this way, we obtain a dispersion of $130^{+50}_{-60}$ km s$^{-1}$. As the dispersion in the northern pointing is 140 km s$^{-1}$, we find a nearly constant velocity dispersion with radius toward the north of Hydra-A.

The central twelve pixels and the outer pixels have bulk velocities of $-10 \pm 20$ km s$^{-1}$ and $-110 \pm 40$ km s$^{-1}$, respectively. The ICM in the central pixels is thus at rest, while the gas in the outer pixels is blueshifted. The implications are discussed in Section \ref{sec:discussion}.

%\FloatBarrier

\begin{figure*}
    \includegraphics[width=\textwidth]{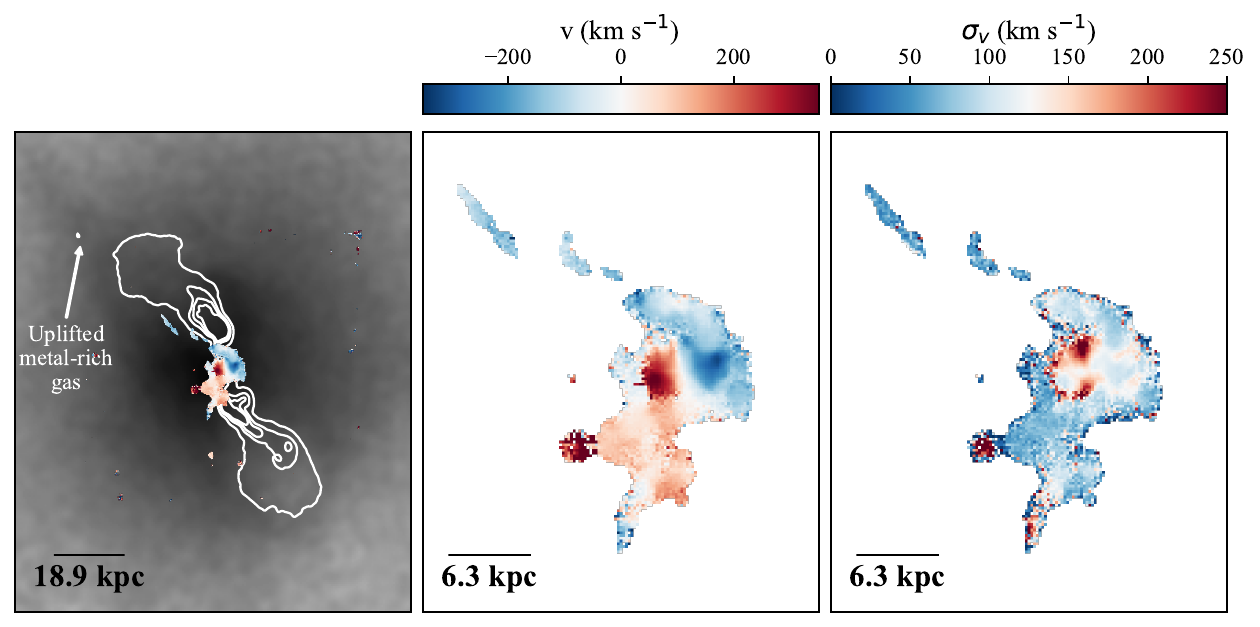}
    \caption{\emph{Left:} Zoomed-in version of Figure 
\ref{fig:central_and_northern_pointings_outlines} (bottom panel), with the 5~GHz EVLA (Project 13B-088) radio lobes and an H$\alpha$ bulk velocity map from MUSE overlaid. \emph{Center:} Bulk velocity map of Hydra-A from MUSE. 
This map is also overlaid on the left panel. \emph{Right:} Velocity 
dispersion map of Hydra-A from MUSE. In the velocity map, note the 
corkscrew-like velocity structure in the gas flowing in the wake of the 
galaxy's radio jets, toward the uplifted metal-rich gas labeled in Figure 
\ref{fig:central_and_northern_pointings_outlines}. Close to the galaxy 
center, there is a large velocity dispersion coincident with the nascent 
radio lobes.}
    \label{fig:MUSE_maps}
\end{figure*}

\section{Discussion} \label{sec:discussion}

\subsection{Energy in the Hydra-A atmosphere}

We can calculate the kinetic energy of the gas using the equation
\begin{equation}
    E_{\textrm{\scriptsize{kinetic}}} = \frac{3}{2} M_{\textrm{\scriptsize{gas}}} \sigma_{\textrm{\scriptsize{1D}}}^2
\end{equation}
\noindent where $M_{\textrm{\scriptsize{gas}}} = 1.78\times 10^{12} M_{\odot}$ is the gas mass in the whole FOV of the central pointing obtained from the radial profiles of \citet{2017ApJ...851...66H}. The estimated kinetic energy in the whole FOV of the central pointing is $1.4 \times 10^{60}$ erg. The summed enthalpies of cavities A, B, and D, which are fully covered by the central pointing (see Figure \ref{fig:central_and_northern_pointings_outlines}), is 
$1.2 \times 10^{60}$ erg \citep{wise_xray_2007}. Additionally, a small fraction of cavity F is covered by the central pointing. The enthalpy of cavity F is $4 \times 10^{60}$ erg, of which a small fraction should contribute to the total gas motion. We thus conclude that the total kinetic energy of the gas motion in the central pointing is comparable to the enthalpy in cavities A, B, and D (and partly F). Gas motion is thus effectively driven by the jet at this scale.

The corresponding masses for the half and quarter footprints of the central 
pointing are $8.9\times 10^{11} M_{\odot}$ and $4.45\times 10^{11} 
M_{\odot}$, assuming spherical symmetry. The kinetic energies in the northern and southern halves are then $1.5 \times 10^{60}$ erg and $2.7 \times 10^{59}$ erg, respectively. The kinetic energies in the eastern and western halves, on the other hand, are $8.6 \times 10^{59}$ erg and $6 \times 10^{59}$ erg, respectively. Finally, the kinetic energies in the northeast, northwest, southeast, and southwest quarters are $9 \times 10^{59}$ erg, $1.9 \times 
10^{59}$ erg, $8.6 \times 10^{58}$ erg, and $1.6 \times 10^{59}$ erg, respectively. 

For the northern pointing, we calculate a line-of-sight-integrated mass of $2.2\times 10^{12} M_{\odot}$. The corresponding kinetic energy is then $1.3 \times 10^{60}$ erg. This energy is roughly an order of magnitude smaller than the enthalpy of cavity E, $\sim 9.9 \times 10^{60}$ erg. We therefore conclude that gas motion at this scale is driven inefficiently by the jet.

\subsection{Asymmetry in velocity dispersion}

Our results indicate an anisotropic velocity dispersion across the cavity system and the radio source. The dispersion in the northern half of the central pointing is $3.9\sigma$ higher than that in the southern half. Such a difference could be due to the northern C and E cavities being more powerful than the southern ones \citep{wise_xray_2007}. The dispersion in the northeast quarter of the footprint is $1.8\sigma$, $2.8\sigma$, and $2.3\sigma$ higher than the dispersions in the northwest, southeast, and southwest quarters, respectively (see Figure \ref{fig:ssm_velocities}). The other quarters all have low ($< 180$ km s$^{-1}$) velocity dispersions and are consistent with each other within $1\sigma$. Given the present data quality, the quarter FOV dispersion anisotropy is tentative. However, if the anisotropy is indeed present, there are two possibilities: unresolved bulk motions and anisotropic turbulence.

The most visually obvious origin for the high dispersion in the northeast quarter is the bright (see Figure \ref{fig:central_and_northern_pointings_outlines}) X-ray structure seen in the Chandra residual image. This feature can also be seen in the smoothed X-ray data of Figure 1 (central panel) in \citet{Kirkpatrick2009} as a coherent, filamentary structure elongated along and aligned with the radio bubbles. They found this region to consist of metal-rich gas probably uplifted by the jet and bubbles. We also see a bulk flow of around $-200$ km s$^{-1}$ near the base and along the direction of this filamentary structure in MUSE, whereas the dispersion is low ($< 100$ km s$^{-1}$; Figure \ref{fig:MUSE_maps}). The MUSE velocity structure is only $\sim$10 kpc in extent, but the X-ray filament is coherent on larger scales of $\sim$50 kpc. Therefore, the data suggest that we could be seeing an unresolved bulk velocity gradient due to such uplift.  Such uplift signatures have been seen in clusters like Virgo \citep{2026A&A...707A.124S}. 

High velocity dispersions ($\sigma_v > 250$ km s$^{-1}$) were previously reported in clusters like Cygnus A \citep{2026ApJ...998..160M} and Virgo \citep{2026ApJ...998..210X}, where unresolved bulk motions are also likely contributing to the large velocity dispersions. The half effective length scale in the core of Hydra-A is $\approx 35$ kpc, according to the definition of \citet{zhuravleva_constraints_2012}. The effective length is approximately the largest possible scale of motion that contributes to the Resolve dispersion.  The full line of sight (LOS) depth, however, is twice this value, i.e., 70 kpc, comparable to the value reported by \citet{rose_xrism_2025}. The effective length in Hydra-A is comparable to the effective length of 20 kpc (40 kpc full LOS depth) in Cygnus A, calculated using the thermodynamic profiles of \citet{majumder_decoding_2024}. Virgo, on the other hand, has a much shorter effective length of 5 kpc \citep{2026ApJ...998..210X} (10 kpc full LOS depth). In all of these cases, the jet is likely an important driver of unresolved bulk motions.

The other possibility is that turbulence in Hydra-A varies azimuthally about the center of the cluster. Turbulence is locally injected by the jet and will diffuse slowly. The diffusion coefficient tends to be small for jet-induced turbulence, and the corresponding diffusion timescale across the core is typically a few Gyr. \citet{2012A&A...544A.103V} simulated jet-induced turbulence in Hydra-A and found a diffusion coefficient of $D = 10^{27}-10^{28}$ cm$^2$ s$^{-1}$. Consequently, the injected turbulence produces strong gas motions near the jet and very weak motions away from it.

If the unresolved bulk motion interpretation is correct, then the true level of turbulence is lower than $\sim$160 km s$^{-1}$ for the full FOV. \citet{rose_xrism_2025} and \citet{2026arXiv260419607M} estimated that turbulence would struggle to balance cooling even under the optimistic assumption that turbulence equals the dispersion. Our results indicate that the true level of turbulence may be as low as $\sim$100 km s$^{-1}$ (as seen in the southern half of the central pointing), and correspondingly, a lower heating rate.

\subsection{Constant mean velocity dispersion along the radio axis} \label{subsec:constant_dispersion}

It is also clear from the results in Section \ref{subsec:radial_profile} that the average dispersion remains flat with increasing radius along the northern direction (see rightmost diagram in Figure \ref{fig:ssm_velocities}). Although gas motions should always be present throughout the ICM due to large-scale processes like mergers and sloshing 
\citep{2012A&A...544A.103V,2013ApJ...762...78Z,2025arXiv251212754B}, one generally expects a rise in gas motion toward the center if the AGN is injecting a significant amount of energy. In Hydra-A, we do not see such a rise despite an energy injection of $2 \times 10^{44}$ erg s$^{-1}$ in the central cavities \citep{wise_xray_2007}. However, this could be due to the large northern cavity, which lies within the FOV of the northern pointing, has a power of $6.8 \times 10^{44}$ erg s$^{-1}$, and is actively being supplied energy by the jet \citep{wise_xray_2007}. The whole region is also near the large-scale cocoon shock front \citep{nulsen_clusterscale_2005,simionescu_large-scale_2009,gitti_chandra_2011} (also see Figure \ref{fig:central_and_northern_pointings_outlines}). We thus need to estimate how much the motion of the shock front affects the velocity dispersion measured by XRISM.

Given the point spread function of XRISM, the shock front is unresolved. Assuming spherical symmetry, the contribution of a small element of an optically thin shock front to the \textit{Resolve} spectrum will be proportional to the area of the element, $dA$, where $dA = r^2 d\Omega$. Here, $r$ is the distance from the center of the shock front and $d\Omega$ is the solid angle subtended by $dA$ at the center of the spherical front. We choose the polar axis of the sphere to be parallel to our line of sight. If the polar angle on the sphere is $\theta$, the velocity along our line of sight of the shocked gas associated with the area element $dA$ is

\begin{equation}
    v_{\textrm{los}} = v_e \textrm{cos}\theta,
\end{equation}
where $v_e$ is the expansion speed of the shocked gas in the cluster frame, and its area is 

\begin{equation}
    dA = r^2 \textrm{sin}\theta d\theta d\phi.
\end{equation}
As $v_{\textrm{los}}$ does not depend on $\phi$, one can integrate over $\phi$, and the distribution function for the projected velocity is then proportional to $dA/|dv_{\textrm{los}}|$, i.e.,

\begin{equation}
    f(v_{\textrm{los}}) \propto \textrm{constant}.
\end{equation}
Therefore, the line-of-sight velocity distribution for a thin spherical shell of the shocked gas would be uniform in the range $[-v_e, v_e]$. The root-mean-square speed along the line of sight for one shell is then simply 

\begin{equation}
    v_{\textrm{ps,RMS}} = \frac{v_e}{\sqrt{3}},
\end{equation}
as expected from the spherical assumption. This is also the dispersion, since the mean along the line of sight for a uniform distribution in the range $[-v_e, v_e]$ is zero.

In general, the shock front is not spherical, so that the shock strength varies over the front. Even for a spherical shock, there is likely to be some spread in the expansion speeds of shells due to the time dependence of the shock strength. The speed of the shock will also be perturbed locally by any pre-existing, small-scale motion in the gas. \citet{nulsen_clusterscale_2005} estimated the Mach number to be in the range $1.2-1.4$.

We now also estimate $v_e$, noting that the velocity jump across a plane 
shock front is

\begin{equation}
    \frac{v_2}{v_1} = \frac{\mathcal{M}^2 (\gamma - 1) + 2}{\mathcal{M}^2(\gamma  + 1)},
\end{equation}
where $v_1$ and $v_2$ are the velocities of the plasma after and before the shock front, respectively, $\mathcal{M}$ is the Mach number of the shock, and $\gamma$ is the ratio of the specific heat capacities. Then,

\begin{equation}
    v_e = v_1 - v_2 = \Big(1 - \frac{v_2}{v_1}\Big)\mathcal{M}c_s = \frac{2c_s}{\gamma + 1} \frac{\mathcal{M}^2 - 1}{\mathcal{M}},
\end{equation}
where $c_s$ is the sound speed of the unshocked gas. This equation assumes that the unshocked gas is at rest.

The gas temperature in the northern pointing (around cavity E) is about 3.5 keV. Therefore, $c_s \approx 965$ km s$^{-1}$ for $\gamma = 5/3$ and $\mu \approx 0.6$. For possible Mach numbers of 1.2, 1.3, and 1.4 across the front, the values of $v_e$ are then approximately 265 km s$^{-1}$, 384 km s$^{-1}$, and 496 km s$^{-1}$, respectively. The corresponding values of $v_{\textrm{ps,RMS}}$ are approximately 153 km s$^{-1}$, 222 km s$^{-1}$, and 286 km s$^{-1}$, respectively. As the dispersion measured from the \textit{Resolve} spectrum of the northern pointing is emission measure weighted along the line of sight, the contribution from denser gas (close to the cluster center) dominates. Since \citet{nulsen_clusterscale_2005} found that the shock is weaker closer to the cluster center, the \textit{Resolve} dispersion is more likely to be representative of a Mach number of 1.2. If this argument holds, then a significant fraction of the dispersion that we see in the northern region could be due to the motion of the large-scale cocoon shock front.

\subsection{Mysterious bulk velocity in the core of Hydra-A}

A statistically significant bulk flow of $-70 \pm 20$ km s$^{-1}$ ($3.5\sigma$ confidence) toward our line of sight exists in the southern half of the central pointing. The motion in the northern half is not statistically significant ($30 \pm 30$ km s$^{-1}$), with the northwest quarter showing the largest value at $100 \pm 40$ km s$^{-1}$ ($2.5\sigma$ confidence). In our discussion, therefore, we will mainly comment on the statistically significant southern bulk flow.

Our results indicate that the southern bulk motion arises away from the center and is concentrated outside the central twelve pixels ($-110 \pm 40$ km s$^{-1}$; $2.75\sigma$ confidence). The outer region spans a distance of $45-110$ kpc from the center. This bulk motion is thus not near the jet. The direction of motion in the southern half (toward us) is also opposite to the motion of the jet itself (away from us; \citealt{Lane2004}), thus suggesting that the bulk motion is not directly related to the jet. The motion is largest in the southwest quarter ($-100 \pm 30$ km s$^{-1}$; $3.3\sigma$ confidence). 

\citet{simionescu_large-scale_2009} reported an apparent offset of $\sim$70 kpc between the center of the cocoon shock ellipse and the center of the radio lobes. They suggested that a large-scale bulk motion is needed to explain such an offset. More recently, \citet{2016A&A...592A.154D} discovered a galaxy group, centered on the galaxy LEDA 87445 \citep{2004AJ....128.1558S}, infalling from the south at a distance of 1.1 Mpc from the cluster core. \citet{2019ApJ...874..112S} interpreted the gas tail behind the galaxy as a slingshot tail, the group passed the cluster center from the northeast at a large impact parameter before receding toward apocenter. Such an interaction could have set off a sloshing motion that produces the bulk motion seen by \textit{Resolve}. Sub-array analysis from \citet{xrism_collaboration_bulk_2025} revealed a similar sloshing motion in the Centaurus cluster. \citet{2026arXiv260814415U} also show a blueshift in the northern regions and a redshift in the southern regions of A2029 (relative to the BCG), which is characteristic of sloshing (also see \citealt{2025PASJ...77S.254S,XRISM_2025_A2029}). In our work, we similarly see an indication of redshift in the northwest quarter ($2.5\sigma$ significance) along with a blueshift in the southwest. Sloshing is therefore a plausible mechanism to explain the bulk velocity. The morphological features of sloshing in the core are difficult to see in the Chandra image, presumably due to the large-scale cocoon shock. This interpretation remains speculative given the limited spatial resolution of the current data. Spatially resolved, high spectral resolution data will be needed in the future to better map the location and dynamics of this bulk flow, and to pinpoint its cause.

\section{Summary}

We analyzed $\sim$116 ks and $\sim$195 ks XRISM observations of the core (central pointing) and northern cavity (northern pointing) of Hydra-A. We performed sub-array analysis of the observations using spatial-spectral mixing. The results can be summarized as follows:

\begin{enumerate}
    \item \textit{Resolve} detects a single-temperature X-ray gas ($kT \sim$ 3.55 keV) around Hydra-A. No statistically compelling evidence for a two-temperature or multi-temperature gas was found. However, a multi-temperature distribution cannot be ruled out, because such a model fit \textit{XMM-Newton} EPIC and RGS data better \citep{simionescu_chemical_2009}, and because \textit{Resolve} is insensitive below 1.7 keV.
    \item The velocity dispersion in the core is $162 \pm 10$ km s$^{-1}$, consistent with previous work by \citet{rose_xrism_2025}. The northern pointing  has a dispersion of $140^{+30}_{-20}$ km s$^{-1}$, i.e., comparable to that of the central pointing.
    \item The dispersions in the eastern and western halves of the central pointing are $180^{+30}_{-40}$ km s$^{-1}$ and $150^{+50}_{-30}$ km s$^{-1}$, respectively. We see a substantial difference ($3.9\sigma$) between the northern and southern halves, with dispersions of  $240^{+40}_{-30}$ km s$^{-1}$ and $100^{+20}_{-30}$ km s$^{-1}$,  respectively. 
    \item Upon dividing the field of view into quarters, we find that most of the dispersion is likely concentrated in the northeast quarter. The velocity dispersions in the northeast, northwest, southeast, and southwest quarters are $260 \pm 50$ km s$^{-1}$, $120^{+60}_{-50}$ km s$^{-1}$, $80 \pm 40$ km s$^{-1}$, and $110^{+40}_{-60}$ km s$^{-1}$, respectively. The northeast dispersion is $1.8\sigma$, $2.8\sigma$, and $2.3\sigma$ higher than the dispersions in the northwest, southeast, and southwest quarters, respectively.
    \item The kinetic energy in the central pointing is $1.4 \times 10^{60}$ erg, consistent with the summed enthalpies of cavities A, B, and D, plus a small fraction of that of cavity F \citep{wise_xray_2007}. Gas motion is thus effectively driven by the jet within the $3' \times 3'$ ($190 \times 190$ kpc) \textit{Resolve} footprint of Hydra-A's center. The northern pointing, on the other hand, has a kinetic energy of $1.3 \times 10^{60}$ erg, roughly an order of magnitude smaller than the enthalpy of cavity E ($4pV \sim 9.9 \times 10^{60}$ erg). Our results suggest that the jet drives gas motion inefficiently at these larger scales, i.e., from $1.5'$ (95 kpc) to $5'$ (317 kpc) toward the north along the jet.
    \item The tentatively higher dispersion in the northeast quarter of the central pointing may be due to the X-ray-bright feature seen with \textit{Chandra} (see Figure \ref{fig:central_and_northern_pointings_outlines}), identified by \citet{Kirkpatrick2009} as a metal-rich region probably produced by uplift by the jet and buoyant cavities. This interpretation would mean that the dispersion in this region is unresolved bulk motion. An alternative interpretation is anisotropic turbulence.
    \item \citet{rose_xrism_2025} and \citet{2026arXiv260419607M} estimated that the turbulent heating rate would struggle to balance cooling in Hydra-A under the optimistic assumption that turbulence equals the velocity dispersion. If the unresolved bulk motion interpretation holds, our results indicate that the true level of turbulence is as low as $\sim$100 km s$^{-1}$, resulting in a correspondingly lower heating rate.
    \item The dispersion of $140^{+30}_{-20}$ km s$^{-1}$ in the northern pointing is likely mostly due to unresolved bulk motion of the cocoon shock front (see Section \ref{subsec:constant_dispersion}). The massive energy injection by cavity E ($4pV \sim 4 \times 10^{60}$ erg; \citealt{wise_xray_2007}) could also be partly responsible.
    \item We detect a statistically significant bulk motion in the southern half of the central pointing ($-70 \pm 20$ km s$^{-1}$; $3.5\sigma$ confidence). The data suggest that the motion is concentrated away from the jet ($45-110$ kpc from the center; $-110 \pm 40$ km s$^{-1}$) and in the southwest quarter ($-100 \pm 30$ km s$^{-1}$). The bulk motion could be due to sloshing from an infalling galaxy group, although this interpretation is speculative for now.
\end{enumerate}

%% Please use the acknowledgment and contribution environments. This will 
%% be anonomyized when the "anonymous" style option is used. 
\begin{acknowledgments}

AM and AS acknowledge support from the Netherlands Organisation for 
Scientific Research (NWO). BRM acknowledges support from the Canadian Space Agency and the Natural Sciences and Engineering Research Council of Canada (NSERC). NW was supported by GACR grant 21-13491X. This research has made use of data obtained from the XRISM data archive maintained by NASA HEASARC and JAXA DARTS. The research has also made use of the \textit{Chandra} data archive provided by the \textit{Chandra} X-ray Center (CXC). Finally, AM acknowledges the use of Claude Opus 4.8 by Anthropic\footnote{\href{https://platform.claude.com/docs/en/about-claude/models/overview}{https://platform.claude.com/docs/en/about-claude/models/overview}} to generate Python code that produced Figures \ref{fig:ssm_regions} and \ref{fig:ssm_velocities}. The accuracy of the code that generated the figures has been verified by the authors. 

\end{acknowledgments}

\section*{Data Availability} \label{sec:data_availability}

The XRISM raw data for the central pointing (ObsID: 201070010) and the northern pointing (ObsID: 201071010) are publicly available on NASA HEASARC and JAXA DARTS. The Chandra data used in this work can be found at ~\dataset[doi.org/10.25574/cdc.643]{https://doi.org/10.25574/cdc.643}. All intermediate data products and code used to obtain the results will be made publicly available at ~\dataset[doi.org/10.5281/zenodo.21364004]{https://doi.org/10.5281/zenodo.21364004} following acceptance of this work in ApJ.

\facilities{XRISM (Resolve), CXO (ACIS), VLT:Yepun (MUSE)}

\software{\texttt{NUMPY} \citep{vdw11}, 
          \texttt{ASTROPY} \citep{astropy:2013,astropy:2018,astropy:2022},
          \texttt{MATPLOTLIB} \citep{hun07},
          \texttt{APLPY} \citep{rob12}}

\appendix
\section{The Sky Background} \label{sec:skybkgsim}

We estimated the effect of the X-ray background (XRB) using a ROSAT all-sky survey (RASS) spectrum. The HEASARC soft X-ray background tool \citep{2019ascl.soft04001S}\footnote{\href{https://heasarc.gsfc.nasa.gov/cgi-bin/Tools/xraybg/xraybg.pl}{https://heasarc.gsfc.nasa.gov/cgi-bin/Tools/xraybg/xraybg.pl}} was used to generate such a spectrum. We extracted the spectrum far from the object to ensure only a sky background contribution to the spectrum. The extraction region was chosen to be an annular region between $1.5R_{200}$ and $1.5R_{200}+1^{\circ}$, where $R_{200} \approx 1190$ kpc \citep{2012PASJ...64...95S}, i.e., 1126 arcsec. The total counts between $0.1-2.4$ keV were 4055.

We fitted a \texttt{hot(cie+pow)+cie} model for Galactic absorption, the absorbed Galactic halo, the absorbed cosmic X-ray background, and the unabsorbed local hot bubble, respectively. The abundance value of each \texttt{cie} component was set to 1. The power-law normalization was fixed to $8.88 \times 10^{-7}$ photons s$^{-1}$ cm$^{-2}$ at 1 keV and the index was fixed to 1.46 (see \citealt{snowden_catalog_2008}). The resulting fit parameters are reported in Table \ref{tab:ROSAT_fit}. The \texttt{hot} parameters were the same as the ones reported in Section \ref{subsec:spec_modelling}.

\begin{figure}
    \centering
	\includegraphics[width=0.5\columnwidth, keepaspectratio]{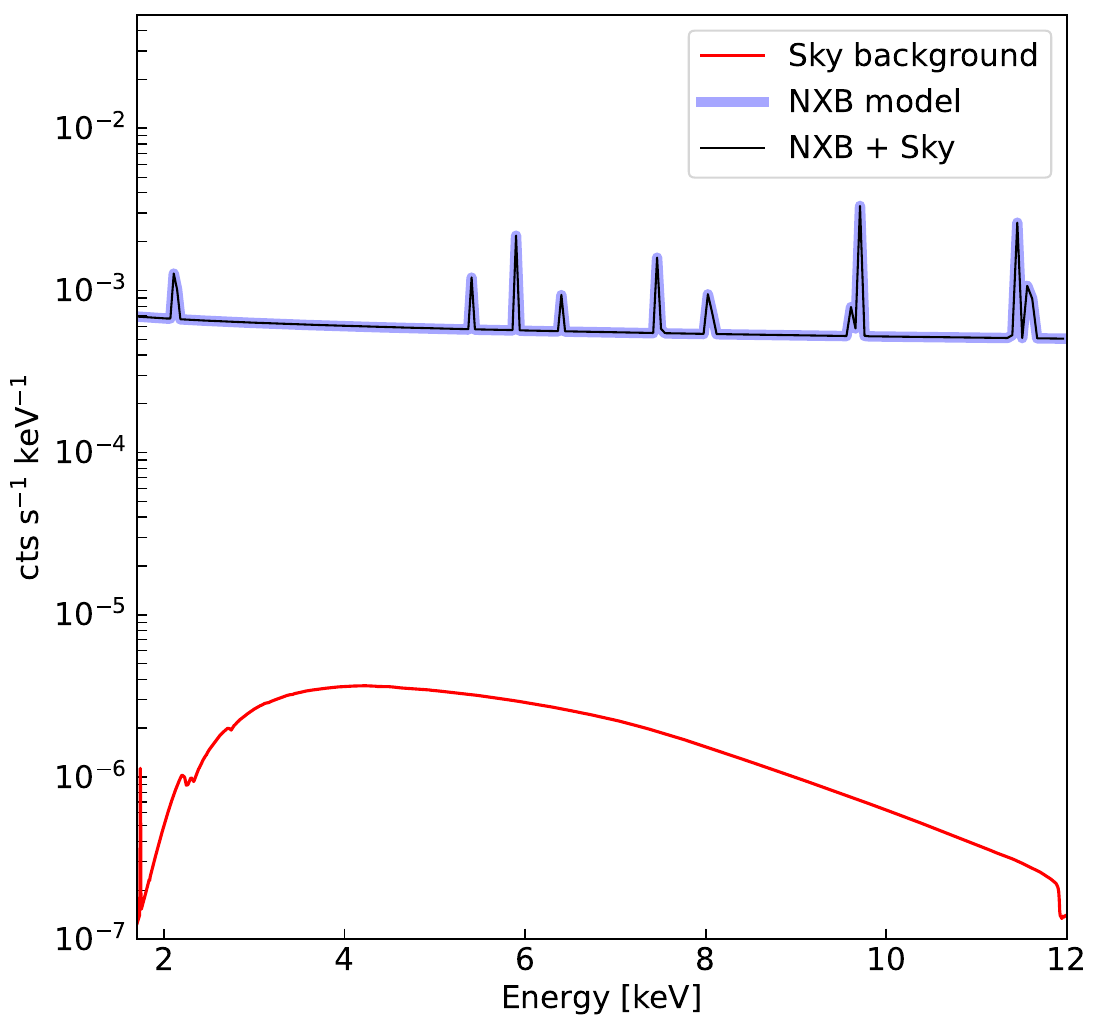}
    \caption{X-ray background spectrum, NXB spectrum, and their sum in the $1.7-12.0$ keV band, simulated for the northern pointing.}
    \label{fig:skybkgsim}
\end{figure}

We used this fitted sky background model and then folded it through the RMF and ARF of the northern pointing. Only the ARF accounting for the photons from the northern pointing region was chosen for demonstration purposes. This folded model is shown together with the NXB spectrum in Figure \ref{fig:skybkgsim}. As can be clearly seen, the sky background is a few orders of magnitude below the NXB and thus the total background (NXB + sky background) is almost entirely due to the NXB. Since the bulk of our northern pointing spectrum lies well over the NXB in the $2-8$ keV band, and the ICM parameter values remain similar irrespective of the chosen band for fitting, we conclude that the inclusion of the sky background will not have a substantial effect on our results. Therefore, the sky background was omitted in all our analysis.

\begin{table}
    \centering
\caption{Best-fit parameters for ROSAT spectrum in the $0.1-2.4$ keV band.}
\label{tab:ROSAT_fit}
    \begin{tabular}{ccc}
    \hline
    \hline
         Components  & Flux ($0.1-2.4$ keV)             & kT \\
                     & (photons s$^{-1}$ keV$^{-1}$)    & (keV) \\
    \hline
         Galactic Halo   &  $1.5_{-0.3}^{+0.4} \times 10^{-2}$ & $0.186_{-0.014}^{+0.017}$\\
         Local Hot Bubble &  $5.18_{-0.13}^{+0.14}\times 10^{-2}$  & $0.104 \pm 0.007$\\
 \hline
    C-stat (Expected C-stat) & \multicolumn{2}{c}{$0.4$ ($7 \pm 4$)}\\
 \hline
 \hline
    \end{tabular}

    \tablecomments{The normalizations are scaled to 1 arcmin$^2$ area.}
\end{table}

\section{Resonance scattering} \label{sec:rsca}

\begin{figure}
    \centering
	\includegraphics[width=0.5\columnwidth, keepaspectratio]{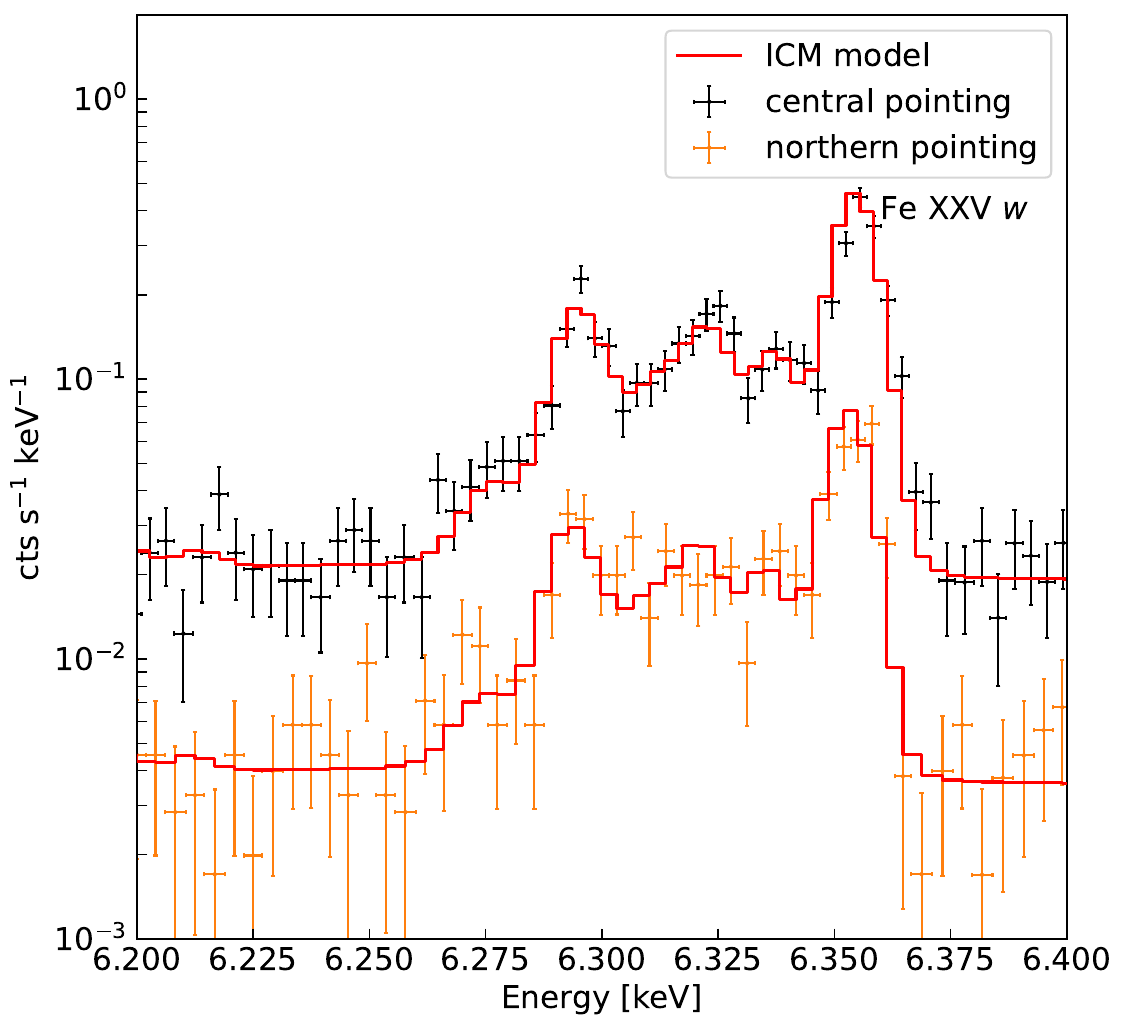}
    \caption{Same as Figure \ref{fig:cn_fit}, but zoomed in between $6.2-6.4$ keV band to show the Fe He $\alpha$ lines.}
    \label{fig:rsca}
\end{figure}

In Figure \ref{fig:rsca}, we show the 1T model fitted to the full-FOV central pointing and full-FOV northern pointing data (see Section \ref{subsec:cenno_ssm}). As can be seen, the \texttt{cie} model, which does not take into account any resonance scattering effect, fits the Fe XXV-$w$ line well. Therefore, there is little evidence for resonance scattering in the core of Hydra A with the present data quality. Despite this, as an additional check, we re-analyzed all our data by ignoring the Fe XXV-$w$ line. The fitted parameter values changed by well under their $1\sigma$ error bounds.

\FloatBarrier

%\bibliography{References_1,References_2}{}
%\bibliographystyle{aasjournalv7}

\end{document}